\documentclass[letterpaper, 10 pt, conference]{ieeeconf}  

\IEEEoverridecommandlockouts                                  
\usepackage{tabularray}
\usepackage{cite}
\usepackage{amsmath,amssymb,amsfonts}
\usepackage[]{footmisc}
\usepackage{xcolor}
\usepackage{graphicx}
\usepackage{textcomp}
\usepackage{psfrag}
\usepackage{epsfig}
\usepackage{subcaption}
\usepackage{float}
\usepackage{xcolor}
\usepackage{bm}
\usepackage{tabularx}
\usepackage{booktabs}

\usepackage[left=1.69cm,right=1.69cm,top=2.01cm,bottom=1.44cm]{geometry}

\newtheorem{myrem}{Remark}

\begin{document}

\title{\LARGE GMPC: Geometric Model Predictive Control for Wheeled Mobile Robot Trajectory Tracking}

\author{Jiawei Tang, Shuang Wu, Bo Lan, Yahui Dong, Yuqiang Jin, Guangjian Tian,\\ Wen-An Zhang,~\IEEEmembership{\textcolor{black}{Senior Member},~IEEE}, and Ling Shi,~\IEEEmembership{Fellow,~IEEE}
	\thanks{The work of Y. Jin and W. Zhang was supported by the National Natural Science Foundation of China under Grant No.  62173305.}
	\thanks{J. Tang, B. Lan, Y. Dong, and L. Shi are with the Department of Electronic
		and Computer Engineering, Hong Kong University of Science and
		Technology, Clear Water Bay, Hong Kong SAR (email: jtangas@connect.ust.hk; blanaa@connect.ust.hk; ydongbb@connect.ust.hk; eesling@ust.hk).}
  \thanks{S. Wu and G. Tian are with the Noah's Ark Lab, Huawei Hong Kong Research Center, Sha Tin, Hong Kong SAR (email:  wushuang.noah@huawei.com; Tian.Guangjian@huawei.com).}
  \thanks{Y. Jin and W. Zhang are with the College of Information Engineering, Zhejiang University of Technology, Hangzhou, China (email: jinyqiang98@gmail.com; wazhang@zjut.edu.cn).}
	}

\maketitle

\begin{abstract}
The configuration of most robotic systems lies in continuous transformation groups. However, in mobile robot trajectory tracking, many recent works still naively utilize optimization methods for elements in vector space without considering the manifold constraint of the robot configuration. In this letter, we propose a geometric model predictive control (MPC) framework for wheeled mobile robot trajectory tracking. We first derive the error dynamics of the wheeled mobile robot trajectory tracking by considering its manifold constraint and kinematic constraint simultaneously. After that, we utilize the relationship between the Lie group and Lie algebra to convexify the tracking control problem, which enables us to solve the problem efficiently. Thanks to the Lie group formulation, our method tracks the trajectory more smoothly than existing nonlinear MPC. Simulations and physical experiments verify the effectiveness of our proposed methods. Our pure Python-based simulation platform is publicly available to benefit further research in the community.
\end{abstract}
\IEEEpeerreviewmaketitle

\section{Introduction}

In recent decades, the wheeled mobile robot (WMR) has been widely applied in many fields, such as autonomous vehicles, intelligent warehouses, and smart agriculture. The widespread adoption of WMRs can be attributed to their remarkable flexibility and efficiency advantages. With technological advancements, WMRs can navigate different terrains, automate complicated tasks, and improve overall productivity. Recent research breakthroughs in robotics further contribute to the growing interest in WMRs \cite{ma-tits-2023, warehouse-ral,caochao-sr}.

The WMR belongs to the class of nonholonomic systems, characterized by a set of nonintegrable first-order differential constraints. These constraints arise from the assumption that wheeled robots move without slipping. Consequently, the nonholonomic constraint of the WMR can be visualized as a situation where the mobile robot cannot undergo lateral translations. According to Brockett's theorem, nonholonomic systems cannot be stabilized solely through smooth time-invariant feedback control laws \cite{261398}. As a result, developing an appropriate controller for achieving trajectory tracking of nonholonomic WMRs is generally a challenging task.

To address the trajectory tracking problem, numerous control schemes have been proposed in the existing literature. Time-varying nonlinear state-feedback controllers have been proposed in \cite{ASTOLFI199637, 261398, 5400088}, and dynamic linear feedback controllers can be found in \cite{fBL, 4812087, PLIEGOJIMENEZ2021109756}. Recently, studies on WMR trajectory tracking with model predictive control (MPC) have appeared. The predictive nature of MPC allows for real-time adaptation and adjustment, making it particularly suitable for dynamic and uncertain environments. The rapid development of computation power benefits the wide dissemination of MPC-based methods \cite{9981038, 10160857,10036044, KHAN2022103903,8967788}.
\begin{figure}[t]
      \centering
	   \begin{subfigure}{0.44\linewidth}
		\includegraphics[width=\linewidth]{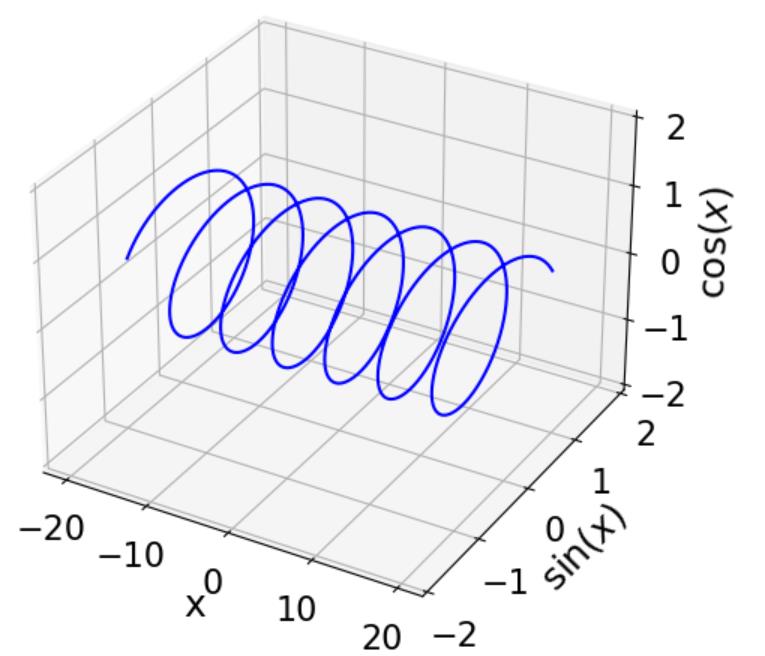}
		\caption{$SO(2)$, Lie group}

	   \end{subfigure}
	   \begin{subfigure}{0.41\linewidth}
		\includegraphics[width=\linewidth]{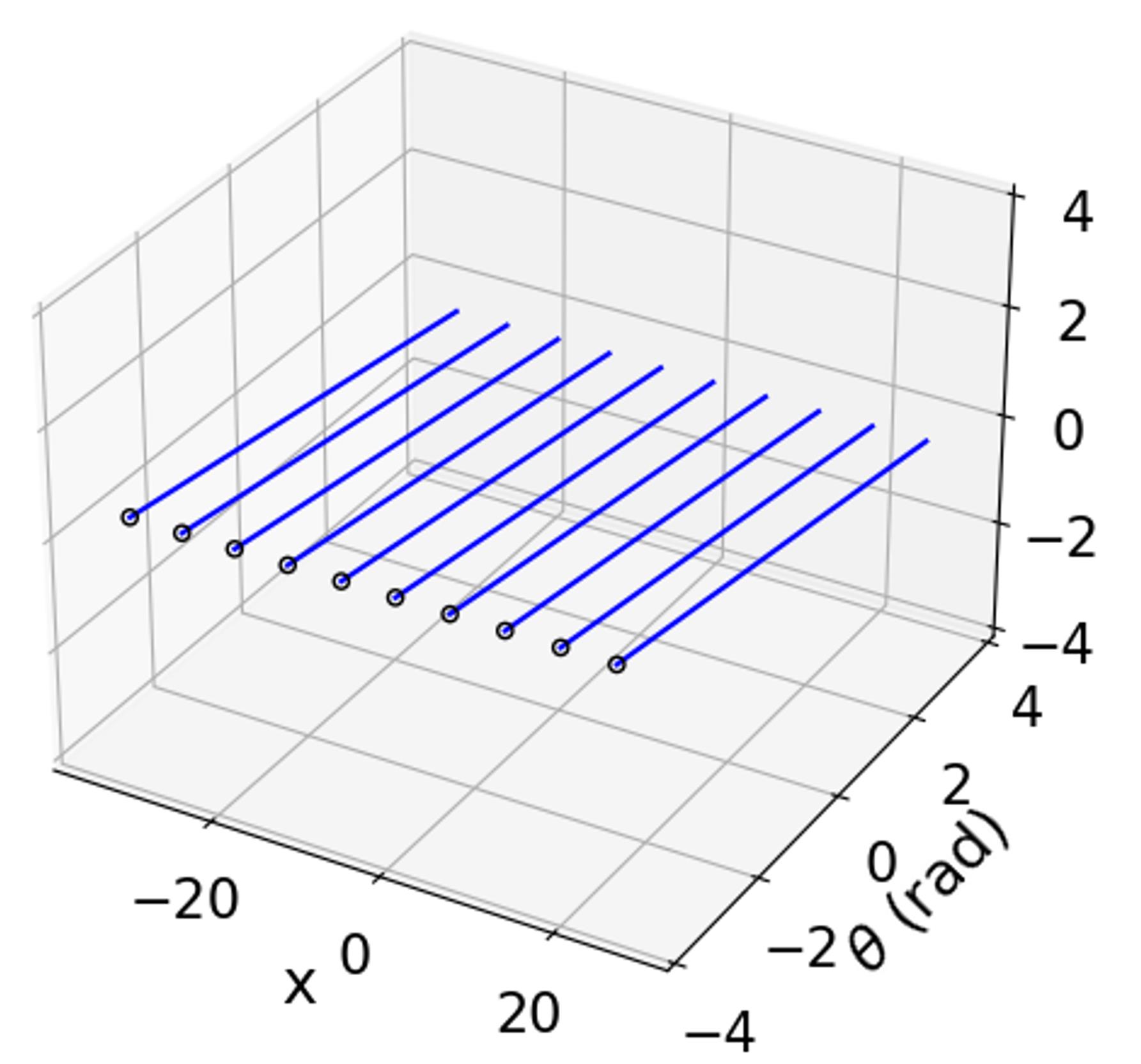}
		\caption{\textcolor{black}{$\theta$, Euler angle}}

	    \end{subfigure}
     \caption{\textcolor{black}{Difference between Lie group and Euler angle representation: The mapping from variable $x$ to Euler angles exhibits discontinuity, whereas the mapping from $x$ to the special orthogonal group $SO(2)$ (an isomorphism of the complex circle group $e^{ix}$) remains continuous.}}
     \label{fig:1}
    \end{figure}
While MPC has shown promising results in modern robotics, a fundamental difficulty lies in effectively incorporating the manifold constraint of the robot configuration into the MPC framework. The configuration of WMR {naturally lies in the continuous transformation group (Lie group) that does not comply with algebraic operations in a vector space \cite{micro-lie-theory}.} For example, the superposition principle does not hold in the matrix Lie group. Hence, theoretical results developed for MPC in vector space cannot be directly extended to the one in the Lie group. {Moreover, when handling the orientation, the singularity issue of the Euler angle and the double-cover issue of the quaternion introduce additional challenges for robotic control and optimization \cite{PWA}.}

The major challenge of Lie group MPC comes from the differential geometry calculus. To deal with the manifold constraints, Jackson et.al. \cite{PWA} developed a modified Newton method for the quaternion, and Alcan et.al. \cite{Alcan2023TrajectoryOO} developed a modified differential dynamic programming for the matrix Lie group. Besides, Lu et.al. \cite{ommpc} developed on-manifold MPC for trajectory tracking. These methods rely on concepts from differential geometry to derive complex derivative calculations and some need to be implemented in customized optimization solvers. Beyond that, many recent works on mobile robots\cite{9981038, 10160857,10036044, KHAN2022103903,8967788} still naively employ optimization methods for elements in vector space without considering the manifold constraints of the robot configuration.

\textcolor{black}{In this letter, we propose a novel geometric MPC (GMPC) framework for trajectory tracking of WMRs using the matrix Lie group. The continuous nature of the matrix Lie group allows for the generation of smoother trajectories compared to the Euler angle-based method (as depicted in Fig. \ref{fig:1}).}  As motivated by the recent research breakthrough on legged robot control\cite{9981282}, we can explore the relationship of the Lie group and the corresponding Lie algebra to convert the state space of MPC from the Lie group to the vector space. Besides, under the framework of the Lie theory, the nonholonomic kinematic constraint of WMR can be easily formulated as a linear mapping from the control input to the velocity of WMR. These advantages enable us to avoid complicated calculations of Lie group derivatives for WMR while handling the WMR's manifold constraint and kinematic constraint simultaneously. Our contributions are multi-fold.


1) We derive the continuous-time error dynamics for the WMR trajectory tracking by considering its manifold constraint and kinematic constraint simultaneously. This derivation enables us to formulate the trajectory tracking as an error-dynamic optimal control problem.

2) {We propose different linearization schemes for the error dynamics to convexification. We show the rationale behind why the proposed linearization scheme is suitable for trajectory tracking and how it helps in the design of convex MPC.}

3) We conduct various simulations and physical experiments with different WMRs and tracking scenarios, which verified the efficiency of our linearization scheme and control framework. Our pure Python-based simulation is open-source to facilitate further research in the community\footnote{https://github.com/Garyandtang/GMPC-Tracking-Control}. 

The remains of this letter are organized as follows. In Section II, some preliminaries on the special Euclidean group and the wheeled mobile robot are presented. In Section III, the main results of the GMPC framework are presented. In Section IV, simulations and physical experiments are provided to evaluate the performance of our method. In Section V, the letter is concluded and some future directions are discussed.

{\noindent{\it Notations}: The main notations used in this letter are as follows.}

\begin{table}[h]
\begin{tabular}{ll}
$\mathbb{R}^{n}$ & \text{$n$-dimensional vector space} \\
 SE(2)& \text{special Euclidean group}\\
 $\mathfrak{se}(2)$& \text{Lie algebra associated with $SE(2)$} \\
 $X \in SE(2)$&\text{state of the rigid body moving in the plane}\\
 $X_d \in SE(2)$&\text{desired state of the rigid body moving in the plane}\\
$\zeta \in \mathbb{R}^3$ & \text{velocity of the rigid body moving in the plane}\\
$\zeta^{\wedge} \in \mathfrak{se}(2)$ & \text{element in Lie algebra}\\
 $u \in \mathbb{R}^2$&\text{control input of the WMR}\\
$\Psi \in SE(2)$&\text{difference between $X_d$ and $X$}\\
$\psi^{\wedge} \in \mathfrak{se}(2)$&\text{Lie algebra associated with $\Psi$}\\
$\dot{X} = \frac{d}{dt}X$&\text{first-order time-derivative of $X$}
\end{tabular}
\end{table}
{\noindent We use $I_{n\times m}$ and $0_{n\times m}$ to represent the $n\times m$ identity matrix and zero matrix, respectively. For notational clarity, the subscript will be dropped if the matrix dimension is clear.}
\section{Preliminaries}
In this section, some useful mathematical results on the special Euclidean group and the wheeled mobile robot are provided, and the problem to be solved is introduced. More details can be found in \cite{micro-lie-theory} for the Lie theory and \cite{fBL} for the wheeled mobile robot.
\subsection{Special Euclidean Group $SE(2)$}
Consider a rigid body in the plane. The position of the rigid body is described by a vector $p = [x,y]^{\top} \in \mathbb{R}^2$, and the orientation is described by $\theta \in S^1$. The orientation can also be represented by a rotation matrix $R \in SO(2)$, where the special orthogonal group $SO(2)$ is defined as
\begin{equation*}
    SO(2) = \{R \in \mathbb{R}^{2 \times 2}~|~R^{\top} R = I, \det R = 1\}.
\end{equation*}
Specifically, the rotation matrix $R \in SO(2)$ is represented as
\begin{equation*}
    R = \begin{bmatrix}
        \cos \theta & -\sin \theta \\ \sin \theta & \cos \theta
    \end{bmatrix}.
\end{equation*}
Hence, the state of the rigid body $X$ can be represented using the homogeneous representation, i.e., 
\begin{equation*}
    X = \begin{bmatrix}
        R & p \\ 0_{1\times 2} & 1
    \end{bmatrix} = \begin{bmatrix}
        \cos \theta & -\sin \theta & x \\ \sin \theta & \cos \theta & y\\  0 & 0 & 1
    \end{bmatrix} \in \mathbb{R}^{3\times 3}.
\end{equation*}
The combination of the set of all $X$ and the operation of matrix multiplication constitute a Lie group known as special Euclidean group $SE(2)$, i.e.,
\begin{equation*}
        SE(2)= \Big\{\begin{bmatrix}
        R & p \\ 0_{1\times 2} & 1
    \end{bmatrix}\in \mathbb{R}^{3 \times 3} ~|~R\in SO(2), ~p \in \mathbb{R}^2\Big\}. 
\end{equation*}
The velocity of the rigid body $\zeta = [v, w]^{\top} \in \mathbb{R}^3$ contains the linear velocity $v = [v_x, v_y]^{\top}\in \mathbb{R}^2$ and the angular velocity $w \in \mathbb{R}$ in the body frame. Under the framework of Lie theory and geometric control, the velocity lies in the tangent space of the Lie group. The space is also known as the Lie algebra $\mathfrak{se}(2)$ of $SE(2)$. The linear map from the velocity $\zeta$ to the element in $\mathfrak{se}(2)$ is denoted as $(\cdot)^{\wedge}$, i.e., 
\begin{align*}
    (\cdot)^{\wedge}&: \mathbb{R}^3 \rightarrow \mathfrak{se}(2);~~~\zeta \rightarrow \zeta^{\wedge} = \begin{bmatrix}
        0 & -w & v_x \\ w & 0 & v_y \\ 0 & 0 & 0
    \end{bmatrix}. \label{wedge}
\end{align*}
The inverse map of $(\cdot)^{\wedge}$ is denoted as $(\cdot)^{\vee}$. Given a continuous time-varying velocity $\zeta(t)$, the rigid body motion on a plane is described as follows
\begin{equation}\label{rigid-body-dyns}
    \dot{X}(t) = X(t) \zeta(t)^{\wedge},
\end{equation}
where $X(t)\in SE(2)$ and $\zeta(t) \in \mathbb{R}^3$. {Notation $\dot{X}(t)$ describes the velocity of the rigid body in the global frame at time $t$}.

\subsection{Exponential and Logarithmic {Map}}
The exponential map $\exp(\cdot)$ provides an exact means of mapping elements from the Lie algebra to the corresponding elements in the Lie group. For elements in $SE(2)$, the exponential map $\exp(\cdot)$ and its inverse map, logarithmic map $\log(\cdot)$, can be expressed as follows
\begin{align*}
    &\operatorname{exp}(\cdot): \mathfrak{se}(2) \rightarrow SE(2), ~\zeta^{\wedge}\rightarrow X = \exp(\zeta^{\wedge}),\\
     &\operatorname{log}(\cdot): SE(2) \rightarrow \mathfrak{se}(2), ~X\rightarrow \zeta^{\wedge} = \log(X) .
\end{align*}
For convenience, we define the capital exponential map and capital logarithmic map to convert vector elements $\zeta \in \mathbb{R}^3$ directly with elements $X \in SE(2)$ as follows
\begin{align}
    &\operatorname{Exp}(\cdot): \mathbb{R}^3 \rightarrow SE(2), ~\zeta\rightarrow X = \operatorname{Exp}(\zeta)\label{EXP-MAP},\\
     &\operatorname{Log}(\cdot): SE(2) \rightarrow \mathbb{R}^3, ~X\rightarrow \zeta = \operatorname{Log}(X) \label{LOG-MAP}.
\end{align}

\subsection{Wheeled Mobile Robot Kinematics}
Consider a nonholonomic wheeled mobile robot that cannot slip in the lateral direction. The corresponding Pfaffian constraint is 
\begin{equation*}
    \dot{x} \sin (\theta) - \dot{y} \cos (\theta) = 0.
\end{equation*}
The kinematic model can be obtained by expressing the entire range of possible velocities, which is shown as follows
\begin{equation}\label{unicycle-model}
    \begin{bmatrix}
        \dot{x} \\ \dot{y} \\ \dot{\theta}
    \end{bmatrix} = \begin{bmatrix}
        \cos \theta & 0 \\ \sin \theta & 0 \\ 0 & 1
    \end{bmatrix}\begin{bmatrix}
        \mu \\ \omega
    \end{bmatrix} = C(\theta) u,
\end{equation}
\textcolor{black}{where $u = [\mu, \omega]^{\top}\in \mathbb{R}^2$ is the control input, including linear velocity $\mu$ and angular velocity $\omega$.} 
\subsection{Trajectory Tracking for the WMR}
Consider the trajectory on special Euclidean group $SE(2)$, we define the motion of the actual state $X(t) \in SE(2)$ and the desired state $X_{d}(t) \in SE(2)$ both as function of time $t$, i.e.,
\begin{equation} \label{traj}
    \dot{X}(t) = X(t) \zeta(t)^{\wedge}, ~~ \dot{X}_d(t) = X_d(t) \zeta_d(t)^{\wedge}.
\end{equation}
\textcolor{black}{Following the group operation to calculate the relative pose from the body frame to reference frame  \cite{iekf}. The error between $X(t)$ and $X_{d}(t)$ is defined as}
\begin{equation} \label{lie-error}
    \Psi(t) = X_{d}(t)^{-1}X(t) \in SE(2).
\end{equation}
For the wheeled mobile robot tracking control, we are interested in the design of control input $u$  such that the error \eqref{lie-error} {can be driven to $I \in SE(2)$} while subject to mobile robot kinematic constraint \eqref{unicycle-model} and control limit constraint. In the following section, we will detail our main ideas for solving the trajectory tracking problem with geometric model predictive control. 
\section{Main Results}
In this section, we introduce our novel control framework to solve the trajectory tracking control problem. We first derive the error dynamics of a rigid body subject to the WMR kinematic constraint and formulate the tracking control problem as a continuous-time optimal control. After that, a convex MPC algorithm is developed based on the Lie theory mechanism to realize real-time performance.
\subsection{{Error-dynamic Optimal Control}}

{Consider the error between the actual trajectory $X(t)$ and the desired trajectory $X_d(t)$. Taking time-derivative on both sides of \eqref{lie-error}, we have
\begin{equation*}
    \begin{aligned}
&\frac{d}{d t} \Psi(t) =\dot{\Psi}(t)=\frac{d}{d t}\left(X_{d}(t)^{-1}\right) X(t)+X_{d}(t)^{-1} \frac{d}{d t} X(t) \\
& =X_{d}(t)^{-1} \frac{d}{d t} X(t)-X_{d}(t)^{-1} \frac{d}{d t}\left(X_{d}(t)\right) X_{d}(t)^{-1} X(t) \\
& =X_{d}(t)^{-1} X(t) \zeta(t)^{\wedge}-X_{d}(t)^{-1} X_{d}(t) \zeta_{d}(t)^{\wedge} X_{d}(t)^{-1} X(t) \\
& =\Psi(t) \zeta(t)^{\wedge}-\zeta_{d}(t)^{\wedge} \Psi(t).
\end{aligned}
\end{equation*}
Hence, we obtain the following error dynamics for a rigid body tracking a reference trajectory in the plane.
\begin{align}
\dot{\Psi}(t)&=\Psi(t) \zeta(t)^{\wedge}-\zeta_{d}(t)^{\wedge} \Psi(t).\label{errorDyn-1}
\end{align}
Recall that the WMR follows the kinematics model described by \eqref{unicycle-model}. The mapping from the control input $u(t)$ to the local velocity $\zeta(t)$ can be obtained with  $C(0)$, i.e.,
\begin{equation} \label{twist-map}
    \zeta(t) = C(0)u(t),
\end{equation}
where $C(0)$ is a constant matrix. Combining \eqref{errorDyn-1} with \eqref{twist-map}, the overall error dynamics of a WMR in $SE(2)$ representation is shown as follows   
\begin{subequations}\label{errorDyn}
    \begin{align}
    \dot{\Psi}(t)&=\Psi(t) \zeta(t)^{\wedge}-\zeta_{d}(t)^{\wedge} \Psi(t),\label{errorDyn-a}\\
    \zeta(t) &= C(0)u(t).\label{errorDyn-b}
    \end{align}
\end{subequations}
\noindent \eqref{errorDyn} implies that given a reference velocity $\zeta_d$, the map from error state $\Psi$ and control $u$ to the generalized velocity $\dot{\Psi}$ is an injective function, and it considers both the rigid body's manifold constraint and the WMR's kinematic constraint simultaneously.} 

Based on the above result on error dynamics, we can formulate the trajectory tracking for the WMR as a continuous-time optimal control.

\noindent\textit{{Problem 1}: (Error-Dynamic Optimal Control)}
\begin{subequations}
\begin{align}
     \min_{{u}(t)} &~~J = \phi(\Psi(t_f))+\int_{0}^{t_f}l(\Psi(\tau), {u}(\tau))d\tau \label{p1_3a}\\
    \text{s.t.} ~~~ &\dot{\Psi}(t)=\Psi(t) \zeta(t)^{\wedge}-\zeta_{d}(t)^{\wedge} \Psi(t), \label{p1_3b}\\     
    &\zeta(t) = C(0)u(t), \\
    & \underline{u} \leq u(t) \leq \overline{u},\\
    &\Psi(0) = \Psi_{\text{init}},\\
    & 0 \leq t \leq t_f,
\end{align}
\end{subequations}
where $\phi(\cdot)$ and  $l(\cdot, \cdot)$ are the terminal cost and the running time cost, respectively, $t_f$ is trajectory duration, $\underline{u}$ and $\overline{u}$ are the lower bound and the upper bound of the control input, respectively. $\Psi_{\text{init}}$ is the initial tracking error.

Note that \textit{Problem 1} is a nonconvex optimization problem since the dynamics constraint \eqref{p1_3b} evolves the Lie group constraint from $SE(2)$ and the cost function \eqref{p1_3a} design should respect the group structure. In what follows, we detail our proposed method to the system dynamics linearization and the cost function design, which benefits the convex MPC formulation.

\subsection{{Convex MPC Formulation}}
Since the exponential map allows us to map the element in Lie algebra to the Lie group. Define $\psi(t)^{\wedge}$ as an element of the Lie algebra $\mathfrak{se}(2)$ corresponding to $\Psi(t)\in SE(2)$. By capital exponential map \eqref{EXP-MAP}, we have
\begin{equation}\label{psi_exp}
    \Psi(t) = \operatorname{Exp}(\psi(t)).
\end{equation}
Taking Taylor expression on \eqref{psi_exp} at the identity, we have
\begin{equation}\label{psi_exp 1}
    \Psi(t) = \operatorname{Exp}(\psi(t)) = I + \sum_{i=1}^{\infty}\frac{1}{i!}(\psi(t)^{\wedge})^i.
\end{equation}
The first-order approximation of \eqref{psi_exp} is as follows
\begin{equation} \label{linear psi}
    \Psi(t) = \operatorname{Exp}(\psi(t)) \approx I + \psi(t)^{\wedge}.
\end{equation}
Taking time-derivative on both sides of \eqref{linear psi}, we have 
\begin{equation}\label{left}
    \dot{\Psi}(t) \approx \frac{d}{dt}(I + {\psi}(t)^{\wedge}) = \dot{\psi}(t)^{\wedge}.
\end{equation}
Substitute \eqref{linear psi} and \eqref{left} into \eqref{errorDyn-a}, we have 
\begin{equation}\label{right}
\begin{aligned}
  \dot{\psi}(t)^{\wedge}  &\approx\Psi(t) \zeta(t)^{\wedge}-\zeta_{d}(t)^{\wedge} \Psi(t)\\
    &\approx(I + \psi(t)^{\wedge})\zeta(t)^{\wedge} - \zeta_{d}(t)^{\wedge} (I + \psi(t)^{\wedge})\\
    &=\zeta(t)^{\wedge} - \zeta_d(t)^{\wedge}+\psi(t)^{\wedge}\zeta(t)^{\wedge}-\zeta_{d}(t)^{\wedge}\psi(t)^{\wedge}.
\end{aligned}
\end{equation}
The nonlinearity in \eqref{right} still persists due to the presence of the high-order term $\psi(t)^{\wedge}\zeta(t)^{\wedge}$. A naive solution to address this issue is to drop this term and approximate \eqref{right} as follows
\begin{equation}\label{linear error-dyn wrong}
\dot{\psi}(t)^{\wedge} \approx \zeta(t)^{\wedge} -\zeta_d(t)^{\wedge}-\zeta_{d}(t)^{\wedge}\psi(t)^{\wedge}.
\end{equation}
\textcolor{black}{However, \eqref{linear error-dyn wrong} is equivalent to \eqref{right} when $\psi(t) = 0$. Nevertheless, $\psi(t)=0$ is attainable only when there is no tracking error. To provide a better approximation, we approximate \eqref{right} as follows}
\begin{equation}\label{linear twist}
\begin{aligned}
&\dot{\psi}(t)^{\wedge} = \zeta(t)^{\wedge} - \zeta_d(t)^{\wedge}+\psi(t)^{\wedge}\zeta(t)^{\wedge}-\zeta_{d}(t)^{\wedge}\psi(t)^{\wedge}\\
&= \zeta(t)^{\wedge} - \zeta_d(t)^{\wedge}+\psi(t)^{\wedge}\zeta_d(t)^{\wedge}-\zeta_{d}(t)^{\wedge}\psi(t)^{\wedge}\\
& +\psi(t)^{\wedge}(\zeta(t)-\zeta_d(t))^{\wedge}\\
&\approx \zeta(t)^{\wedge} - \zeta_d(t)^{\wedge}+\psi(t)^{\wedge}\zeta_d(t)^{\wedge}-\zeta_{d}(t)^{\wedge}\psi(t)^{\wedge}.
\end{aligned}
\end{equation}
 \textcolor{black}{Note that \eqref{linear twist} is equivalent to \eqref{right} when $\zeta(t) = \zeta_d(t)$. Since we aim to minimize the different between $\zeta(t)$ and $\zeta_d(t)$, the the high-order term $\psi(t)^{\wedge}(\zeta(t)-\zeta_d(t))^{\wedge}$ we drop in the last approximation of \eqref{linear twist} will tend to be small. This indicates that \eqref{linear twist} is a better approximation to \eqref{right} than \eqref{linear error-dyn wrong}. We will conduct a numerical comparison of these two linearization schemes in Section IV.}

Note that the operation $(\cdot)^{\wedge}$  is linear. Therefore, \eqref{linear twist} is a linearized error dynamics for the WMR. Take $(\cdot)^{\vee}$ operation on both sides on \eqref{linear twist}, we have
\begin{equation}\label{linear twist 2}
    \dot{\psi}(t) = \zeta(t) - \zeta_d(t)+\operatorname{adm}(\zeta_d(t))\psi(t),
\end{equation}
where $\operatorname{adm}(\zeta_d(t))$ is expressed as follows
\begin{equation*}
    \operatorname{adm}(\zeta_d(t)) = \begin{bmatrix}
        0 &  w_d(t) & -v_{y,d}(t)\\ -w_d(t) & 0 & v_{x,d}(t)\\ 0 & 0 & 0
    \end{bmatrix}.
\end{equation*}
Combining \eqref{linear twist 2} with \eqref{errorDyn-b}, the overall linearized error dynamics of a wheeled mobile robot is 
\begin{equation}
    \begin{aligned}
        \dot{\psi}(t) =A(t)\psi(t) + B(t)u(t) + c(t), \label{final-linear-dyn}
    \end{aligned}
\end{equation}
where
\begin{equation*}
    A(t):= \operatorname{adm}(\zeta_d(t)), ~~ B(t) := C(0), ~~ c(t) := - \zeta_d(t).
\end{equation*}

  Different from \eqref{errorDyn}, the state and control input in \eqref{final-linear-dyn} are in vector space. The above linearization process allows us to utilize the algebraic operations of vector space and further enables us to use existing off-the-shelf solvers to solve optimal control problems.  

As we convert the problem to vector space, we can directly use the weighted Euclidean norm to penalize the state and control input. Thus, we define the cost function for tracking control as follows
\begin{equation*}
    \begin{aligned}
        J = \psi(t_f)^{\top}Q_f\psi(t_f)+\int_0^{t_f}\psi(\tau)^{\top}Q\psi(\tau)+ \hat{u}(\tau)^{\top}H\hat{u}(\tau)d\tau,
    \end{aligned}
\end{equation*}
where $Q_f\in \mathbb{R}^{3 \times 3}$, $Q\in \mathbb{R}^{3 \times 3}$ and $H\in \mathbb{R}^{2 \times 2}$ are the final state penalty weight matrix, intermediate state penalty weight matrix, and the control penalty weight matrix, respectively. $\hat{u}(t) = u(t) - u_d(t)$, and $u_d(t)$ is the control input of the reference trajectory.

Based on the above results on error dynamics linearization and objective function design, we have the following linear quadratic optimal control problem for trajectory tracking.

\noindent\textit{{Problem 2}: (Linear Quadratic Optimal Control)}
\begin{align*}
     \min_{{u}(t)} &~\psi(t_f)^{\top}Q_f\psi(t_f)+\int_0^{t_f}\psi(\tau)^{\top}Q\psi(\tau)+ \hat{u}(\tau)^{\top}H\hat{u}(\tau)d\tau\\
    \text{s.t.} &~\dot{\psi}(t) =A(t)\psi(t) + B(t)u(t) + c(t), \\ 
    &~\underline{u} \leq u(t) \leq \overline{u},\\
    &~\psi(0) = \psi_{\text{init}},\\
    & ~0 \leq t \leq t_f,
\end{align*}
where $\psi_{\text{init}} = \operatorname{Log}(\Psi_{\text{init}}$).

Note that \textit{Problem 2} is a continuous-time optimal control problem. To make it solvable by off-the-shelf optimization solvers, we utilize the direct transcription method \cite{betts2010practical} to discrete the continuous-time functions to some sequences of $N+1$ real numbers, i.e., for $0 \leq t \leq t_f$,
 \begin{align*}
t &\rightarrow t_0, \ldots, t_k, \ldots,t_N,\\
{\psi}(t) &\rightarrow {\psi}_0 , \ldots, {\psi}_k, \ldots,{\psi}_N,\\
{u}(t) &\rightarrow {u}_0 , \ldots, {u}_k, \ldots, {u}_{N},
\end{align*}     
where $\psi_k$ and $u_k$ are the approximations to $\psi(t_k)$, $u(t_k)$ at $t = t_k$, respectively. Therefore, the finite-dimensional convex MPC tracking control is formulated as follows

\noindent\textit{{Problem 3}: (Convex MPC {Tracking Control})}
\begin{subequations}
\begin{align}
    \min_{\{u_k\}_{k=0}^{T-1}} &~\psi_T^{\top}Q_f\psi_T+\sum_{k=0}^{T-1}\psi_k^{\top}Q_k\psi_k+ \hat{u}_k^{\top}H_k\hat{u}_k\\
    \text{s.t.} &~{\psi}_{k+1} =A_k\psi_k + B_ku_k + c_k, \\
    &~\underline{u} \leq u_k \leq \overline{u},\\
    &~\psi_0 = \psi_{\text{init}},\\
    & ~k = 0, 1, \dots, N-1,
\end{align}
\end{subequations}
\noindent where $Q_k$, $H_k$, $A_k$, $B_k$, $c_k$ can be obtained through numerical integration and $T \leq N$ is the prediction horizon. Since all equality constraints are linear and the objective is quadratic, \textit{Problem 3} can be easily solved by off-the-shelf quadratic programming (QP) solvers, such as OSQP\cite{osqp} and qpOASES \cite{Ferreau2014}.
\begin{myrem}
    In our implementation, we use the Euler method for numerical integration, i.e., 
    \begin{align*}
        A_k &= I + A(t_k) \Delta t,~ B_k = B(t_k)\Delta t, ~c_k = c(t_k)\Delta t,\\
        Q_k &= Q \Delta t,~ H_k = H \Delta t,~ \Delta t = t_k - t_{k-1}.
    \end{align*}
    Different numerical integration methods, such as the Runge-Kutta and Trapezoidal methods\cite{betts2010practical}, are also applicable in the MPC formulation. 
\end{myrem}

\section{Experiments}

In this section, we conduct simulations and physical experiments to evaluate the performance of the proposed control framework on two types of wheeled mobile robots.
\subsection{Simulation Experiments}
\begin{figure}[t]
      \centering
	   \begin{subfigure}{0.42\linewidth}
		\includegraphics[width=\linewidth]{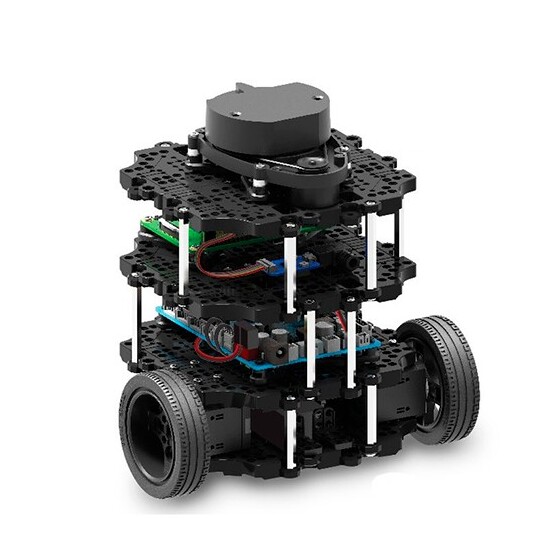}
		\caption{Turtlebot 3}
		\label{fig:turtle}
	   \end{subfigure}
	   \begin{subfigure}{0.42\linewidth}
		\includegraphics[width=\linewidth]{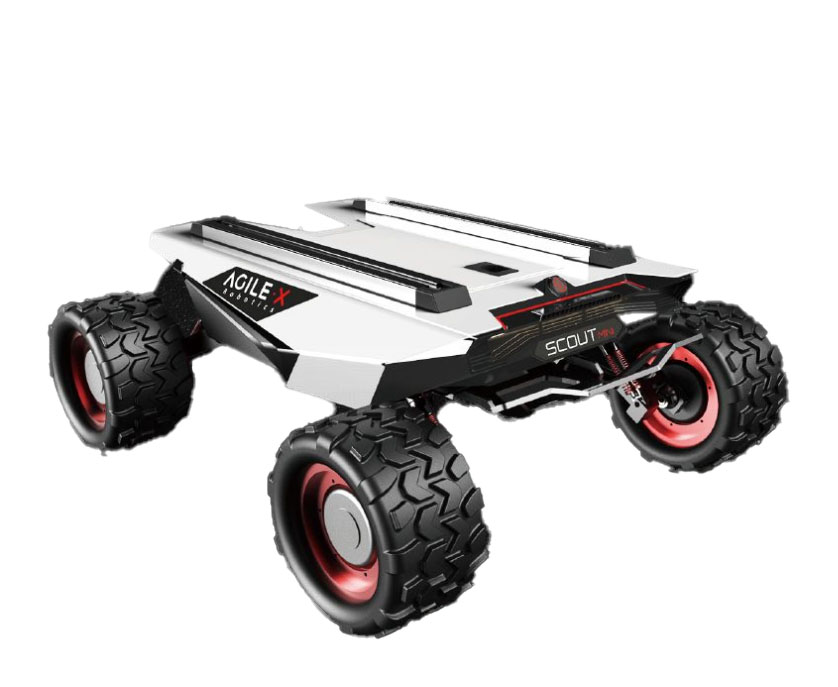}
		\caption{\textcolor{black}{Scout mini}}
		\label{fig:scout}
	    \end{subfigure}
\caption{Wheeled mobile robot platforms.}
\end{figure}

Our simulation platform is created in PyBullet\cite{pybullet} with the symbolic framework CasADi\cite{andersson2019casadi}. The Lie-group-related calculation is implemented based on Manif\cite{Manif}. We consider two types of wheeled mobile robots, i.e., the two-wheel-drive Turtlebot 3\footnote{https://emanual.robotis.com/docs/en/platform/turtlebot3/overview/} (Fig. \ref{fig:turtle}) and the four-wheel-drive Scout mini\footnote{\textcolor{black}{https://global.agilex.ai/more/download/11}} (Fig. \ref{fig:scout}). The system parameters of these two WMRs set in the simulation are summarized in TABLE \ref{tab:system param}.

\begin{figure}[t]
      \centering
	   \begin{subfigure}{0.44\linewidth}
		\includegraphics[width=\linewidth]{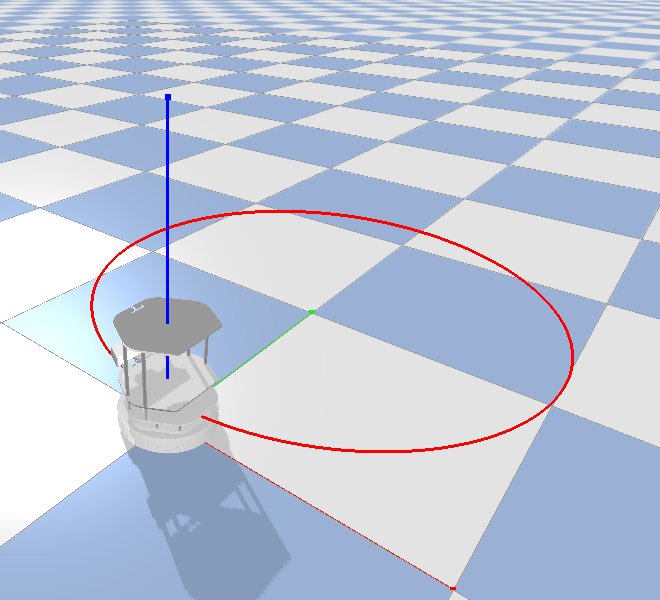}
		\caption{{Circular trajectory}}
		\label{fig:subfig1}
	   \end{subfigure}
	   \begin{subfigure}{0.44\linewidth}
		\includegraphics[width=\linewidth]{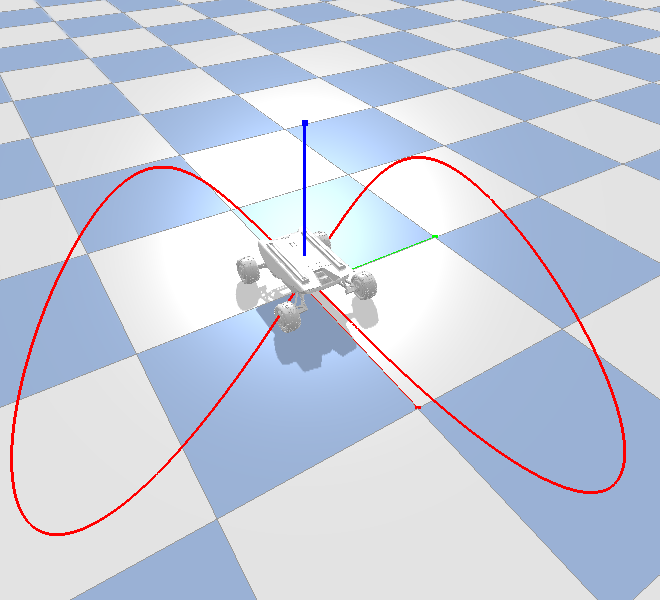}
		\caption{{Butterfly-shaped trajectory}}
		\label{fig:subfig2}
	    \end{subfigure}
     \caption{Trajectory tracking scenarios.}
     \label{fig:trajectory types}
\end{figure}

We consider two trajectory tracking scenarios as shown in Fig. \ref{fig:trajectory types}: (a) a circular trajectory with constant control input $u_d$, and (b) a butterfly-shaped trajectory with time-varying control input. The trajectories are integrated by $u_d$ from the origin through the forward Euler method. To evaluate the robustness performance, we randomly initialize the start pose of WMRs around the origin.

\begin{table}[t]
    \caption{System parameters of two WMR platforms}
    \label{table_example}
    \begin{center}
    \resizebox{.5\textwidth}{!}{
    	\renewcommand\arraystretch{1.5}
        \begin{tabular}{c|ccccc}
        \bottomrule \hline
            Platform & Min $\mu$ & Max $\mu$ & Min $\omega$ & Max $\omega$ & Ctrl. Freq.  \\
        \hline
            Turtlebot 3 & \textcolor{black}{-0.22 m/s} & \textcolor{black}{0.22 m/s} & \textcolor{black}{-2.84 rad/s} & \textcolor{black}{2.84 rad/s} & 50 Hz \\
            Scout mini& \textcolor{black}{-3 m/s} & \textcolor{black}{3 m/s} & \textcolor{black}{-2.523 rad/s} & \textcolor{black}{2.523 rad/s} & 50 Hz\\
        \hline \toprule 
        \end{tabular}
    }
    \end{center}
    \label{tab:system param}
\end{table}
We compare our method (\textit{GMPC}) with two baseline control frameworks: (a) a nonlinear model predictive control (\textit{NMPC}) framework with orientation represented by Euler angle proposed in \cite{10160857}\footnote{As stated in the introduction, many recent works also rely on a similar Euler-based formulation in nonlinear MPC implementation\cite{9981038, 10160857,10036044, KHAN2022103903,8967788}.} and (b) a feedback linearization tracking controller (\textit{FBC}) proposed in \cite{fBL}. We also compare different system dynamics linearization schemes in our \textit{GMPC}, which are described in \eqref{linear error-dyn wrong} and \eqref{linear twist}. Since the performance of tracking controllers is highly dependent on the parameter tuning, we carefully turn the penalty matrices $Q$, $Q_f$, and $H$ of \textit{GMPC} and \textit{NMPC} and the feedback gain of \textit{FBC}. In addition, We set the prediction horizon $T = 10$ for the MPC-based controllers. The GMPC is implemented with qpOASES\cite{Ferreau2014}, and the NMPC is implemented with IPOPT\cite{wachter2006implementation}. 

\begin{figure}[t]
      \centering
	   \begin{subfigure}{0.48\linewidth}
		\includegraphics[width=\linewidth]{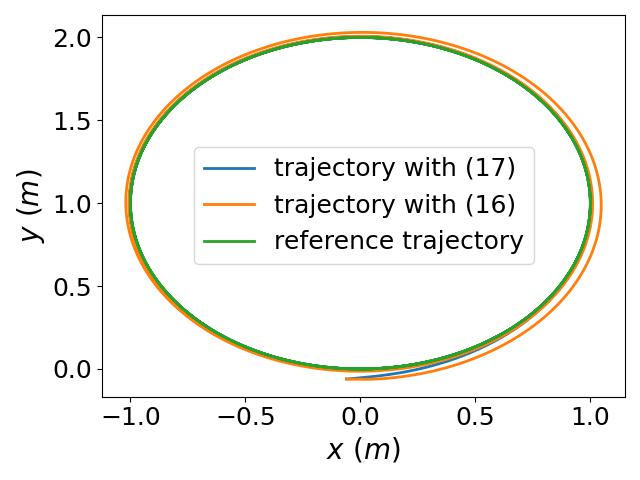}
		\caption{\textcolor{black}{Trajectories}}
		\label{fig:subfig11}
	   \end{subfigure}
	   \begin{subfigure}{0.48\linewidth}
		\includegraphics[width=\linewidth]{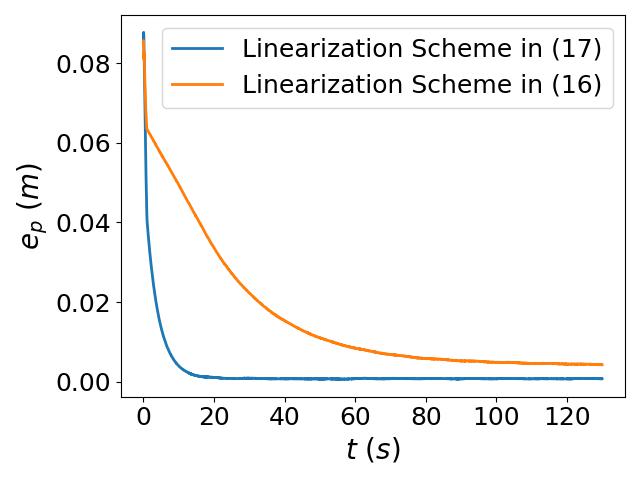}
		\caption{\textcolor{black}{Position errors}}
		\label{fig:subfig22}
	    \end{subfigure}
     \caption{\textcolor{black}{Circular trajectory tracking with Turtlebot 3 using different linearization schemes. The initial position of the robot is $[-0.06, -0.06]^{\top}$, and the initial orientation is $0$.}}
     \label{fig: linearization comparison}
\end{figure}
We follow \cite{Lee2019IntroductionTR} to evaluate the tracking error between the actual trajectory and the reference trajectory. Specially,  the position error $e_p(t)$ is obtained by Euclidean norm, i.e., 
\begin{equation*}
    e_p(t) = \|p(t) - p_d(t)\|_2,
\end{equation*}
where $p(t)$ and $p_d(t)$ are the actual and the reference position, respectively. The orientation error is obtained by Riemannian metric, i.e.,
\begin{equation*}
    e_{R}(t) = \|\operatorname{Log}(R_d(t)^{-1}R(t))\|_2,
\end{equation*}
where $R(t)$ and $R_d(t)$ are the actual and the reference rotation matrix, respectively.

\begin{figure*}[htb]
    \centering
    \begin{subfigure}{0.33\textwidth}
        \includegraphics[width=\textwidth]{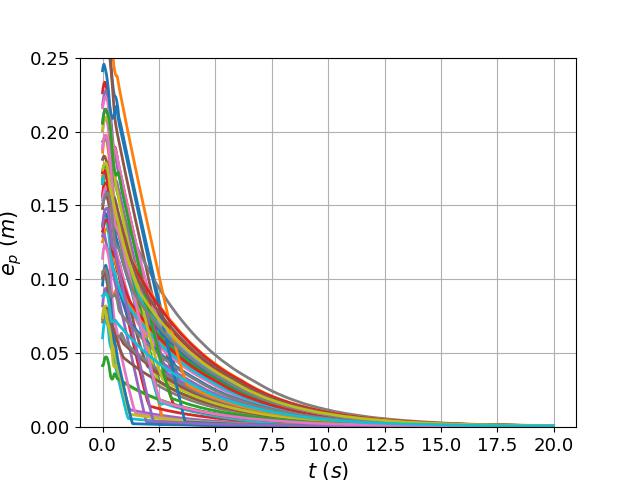}
        \caption{Position error $e_p(t)$ (GMPC)}
        \label{fig:BigMap1_a}
    \end{subfigure}
    \begin{subfigure}{0.33\textwidth}
        \includegraphics[width=\textwidth]{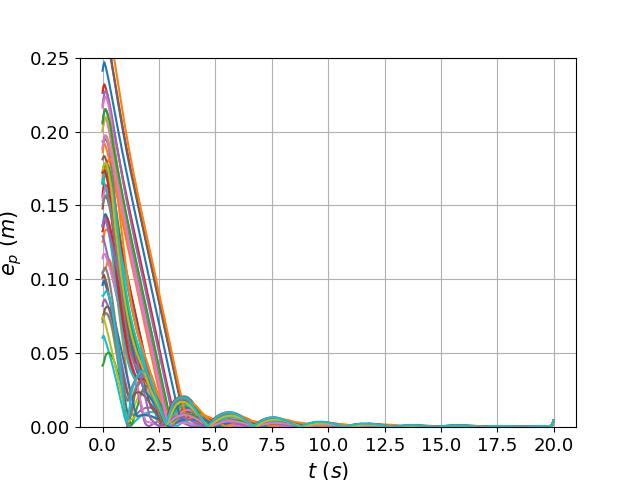}
        \caption{Position error $e_p(t)$ (NMPC)}
        \label{fig:BigMap1_b}
    \end{subfigure}%
    \begin{subfigure}{0.33\textwidth}
        \includegraphics[width=\textwidth]{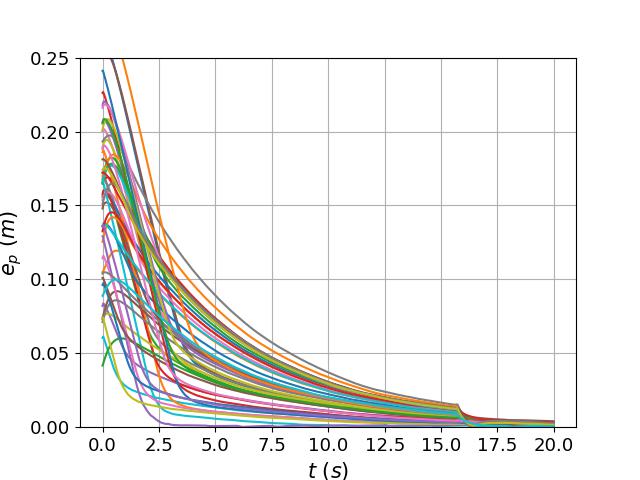}
        \caption{Position error $e_p(t)$ (FBC)}
        \label{fig:BigMap1_c}
    \end{subfigure}%
    \quad
    \begin{subfigure}{0.33\textwidth}
        \includegraphics[width=\textwidth]{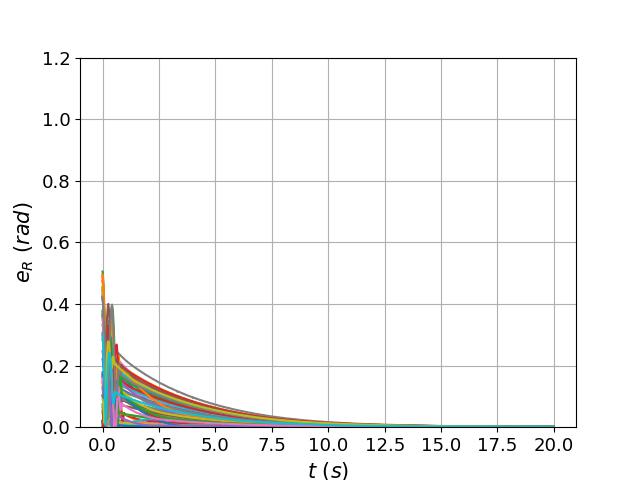}
        \caption{Orientation error $e_R(t)$ (GMPC)}
        \label{fig:BigMap1_a}
    \end{subfigure}
    \begin{subfigure}{0.33\textwidth}
        \includegraphics[width=\textwidth]{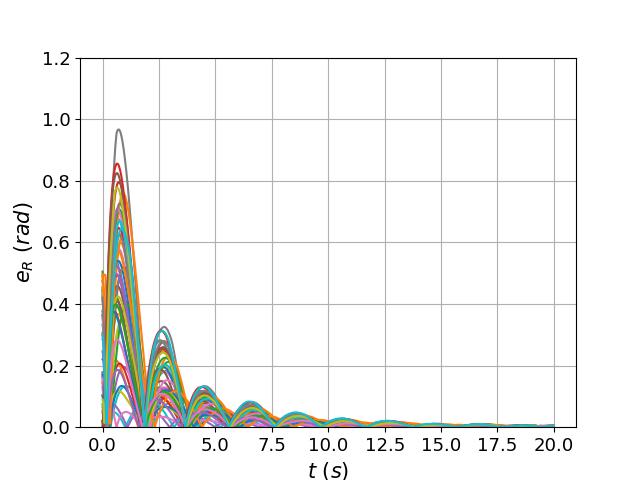}
        \caption{Orientation error $e_R(t)$ (NMPC)}
        \label{fig:BigMap1_b}
    \end{subfigure}%
    \begin{subfigure}{0.33\textwidth}
        \includegraphics[width=\textwidth]{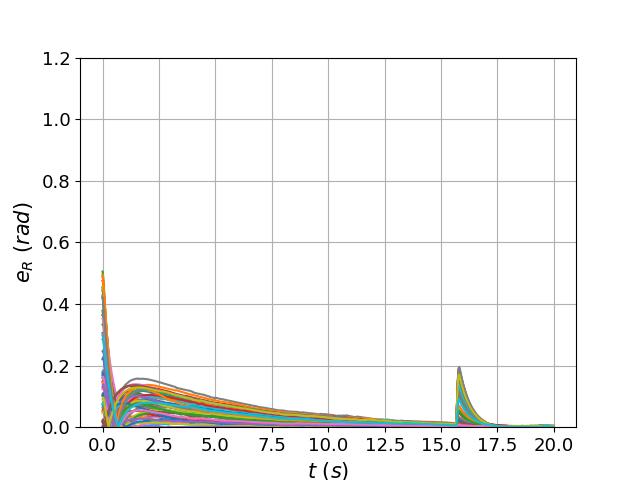}
        \caption{Orientation error $e_R(t)$ (FBC)}
        \label{fig:BigMap1_c}
    \end{subfigure}%
   
    \caption{\textcolor{black}{Monte Carlo tests of tracking a circular trajectory with Turtlebot 3. The initial position of the robot is randomly selected between $[-0.2, 0]^{\top}$ and $[-0.2, 0]^{\top}$, and the initial orientation is randomly selected between $[-\frac{\pi}{6},0]$.}}
    \label{fig:MC-TEST}
\end{figure*}

\begin{figure*}[htb]
    \centering
    \begin{subfigure}{0.2\textwidth}
        \includegraphics[width=\textwidth]{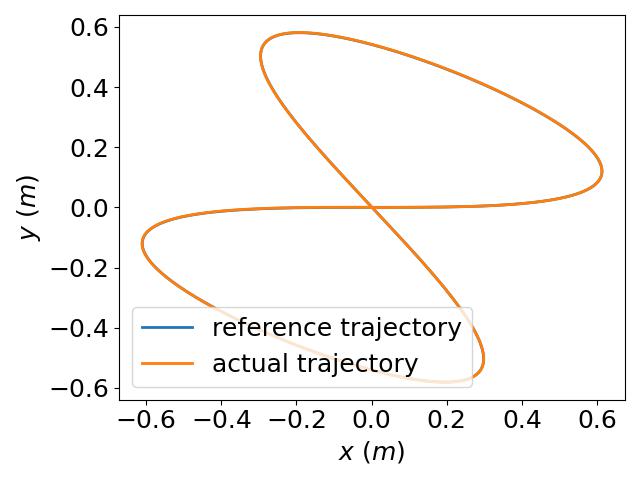}
        \caption{Trajectory (GMPC)}
       
    \end{subfigure}
    \begin{subfigure}{0.2\textwidth}
        \includegraphics[width=\textwidth]{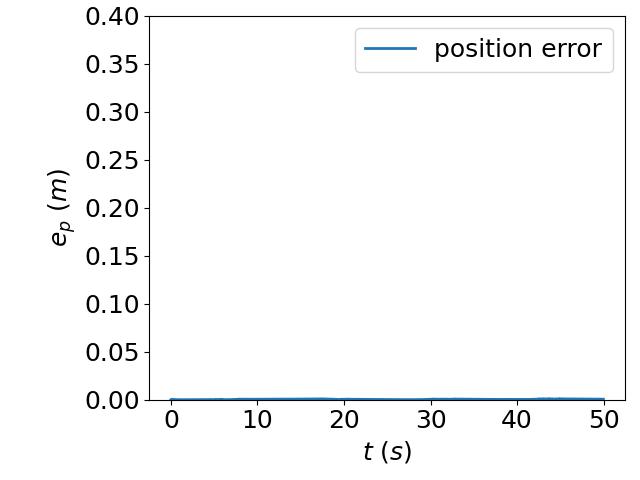}
        \caption{Position error}
       
    \end{subfigure}%
    \begin{subfigure}{0.2\textwidth}
        \includegraphics[width=\textwidth]{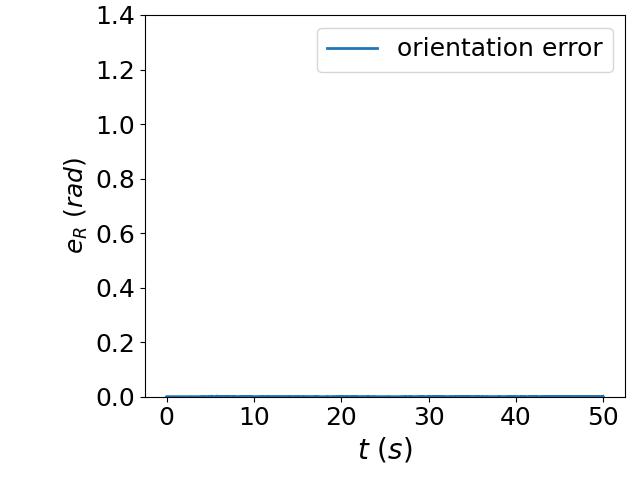}
        \caption{Orientation error}
      
    \end{subfigure}%
    \begin{subfigure}{0.2\textwidth}
        \includegraphics[width=\textwidth]{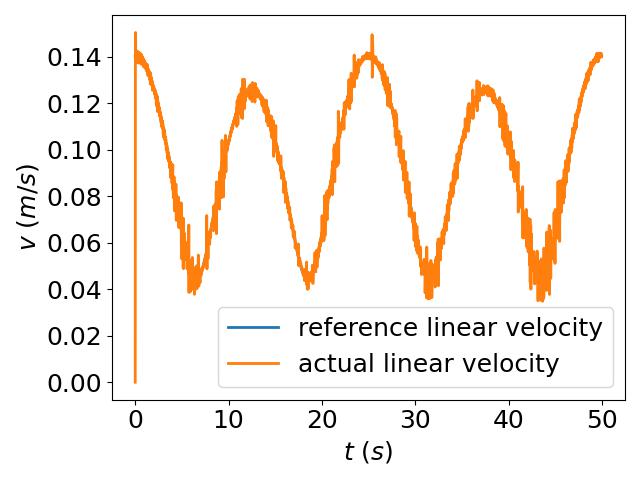}
        \caption{Linear velocity}
      
    \end{subfigure}%
    \begin{subfigure}{0.2\textwidth}
        \includegraphics[width=\textwidth]{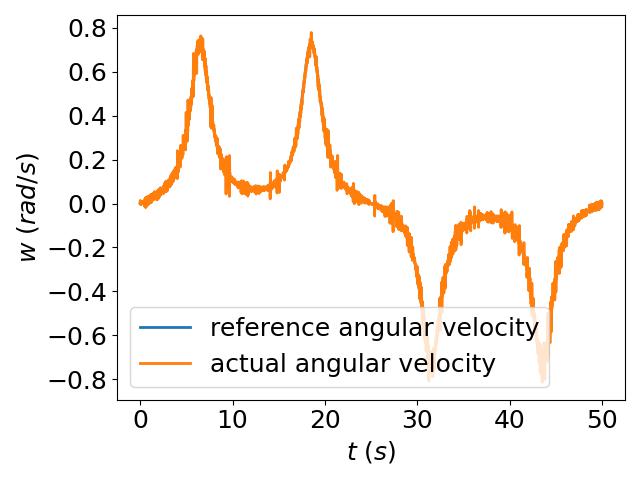}
        \caption{Angular velocity}
      
    \end{subfigure}%

    \quad
    \begin{subfigure}{0.2\textwidth}
        \includegraphics[width=\textwidth]{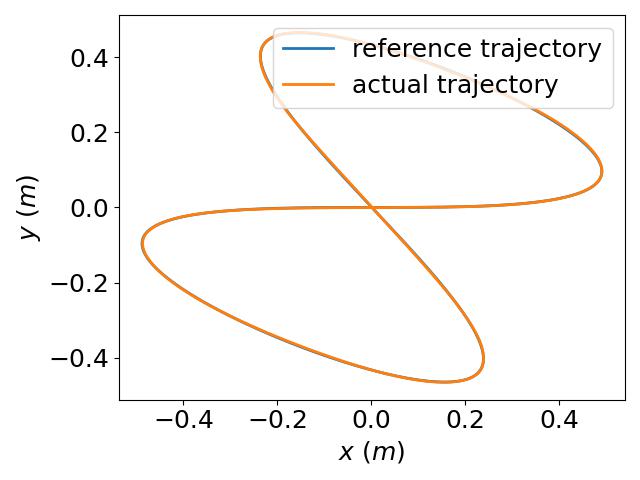}
        \caption{Trajectory (NMPC)}
       
    \end{subfigure}
    \begin{subfigure}{0.2\textwidth}
        \includegraphics[width=\textwidth]{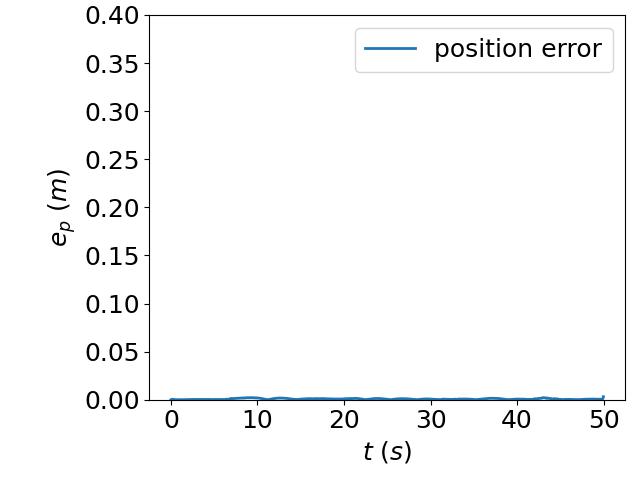}
        \caption{Position error}
       
    \end{subfigure}%
    \begin{subfigure}{0.2\textwidth}
        \includegraphics[width=\textwidth]{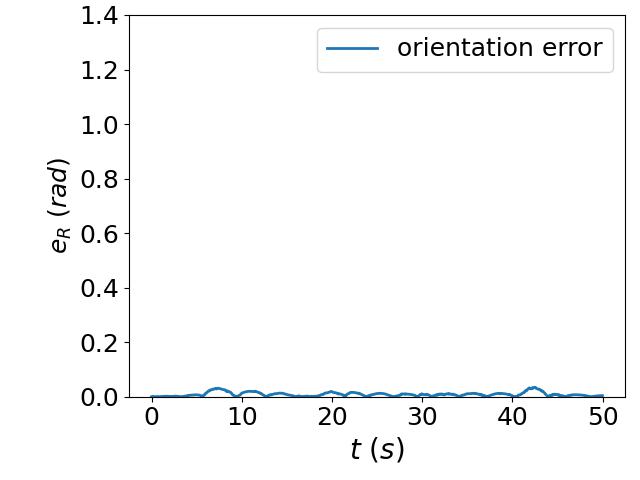}
        \caption{Orientation error}
      
    \end{subfigure}%
    \begin{subfigure}{0.2\textwidth}
        \includegraphics[width=\textwidth]{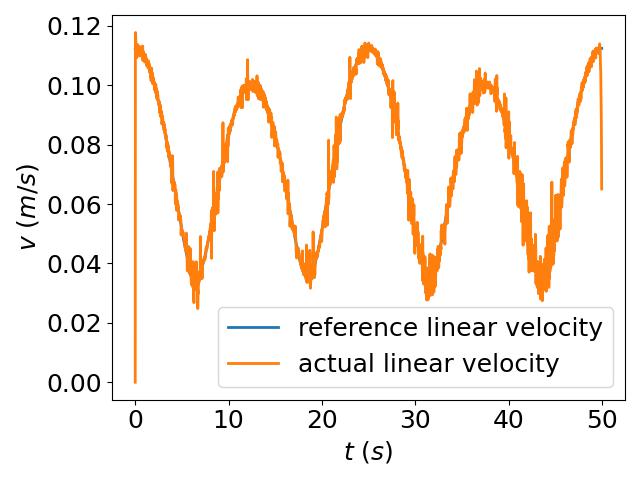}
        \caption{Linear velocity}
      
    \end{subfigure}%
    \begin{subfigure}{0.2\textwidth}
        \includegraphics[width=\textwidth]{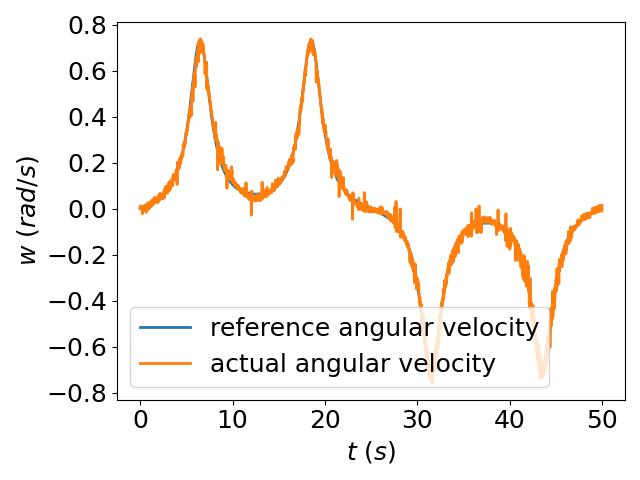}
        \caption{Angular velocity}
      
    \end{subfigure}%

    \quad
    \begin{subfigure}{0.2\textwidth}
        \includegraphics[width=\textwidth]{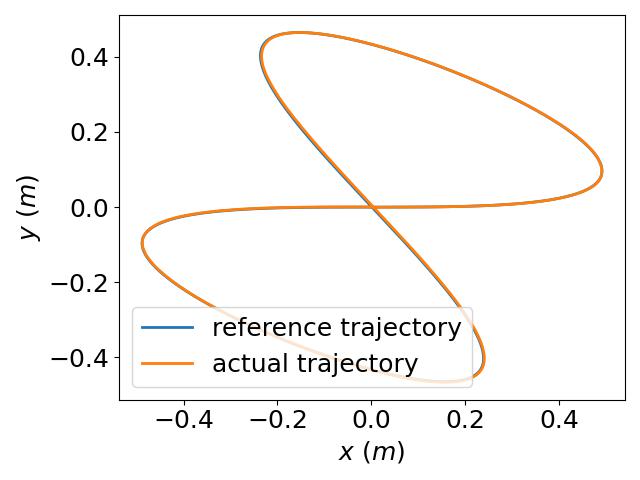}
        \caption{Trajectory (FBC)}
       
    \end{subfigure}
    \begin{subfigure}{0.2\textwidth}
        \includegraphics[width=\textwidth]{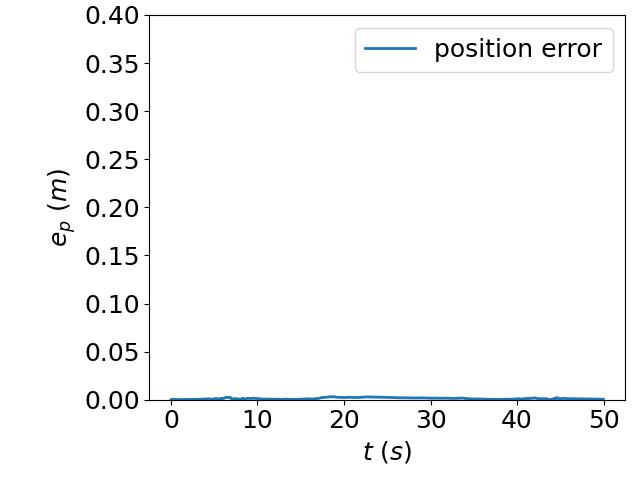}
        \caption{Position error}
       
    \end{subfigure}%
    \begin{subfigure}{0.2\textwidth}
        \includegraphics[width=\textwidth]{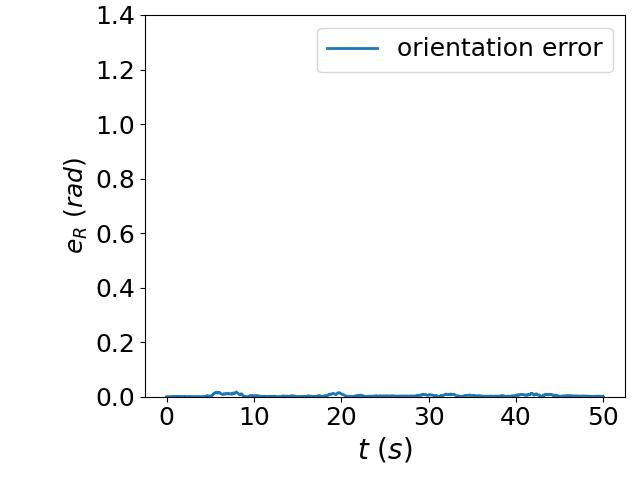}
        \caption{Orientation error}
      
    \end{subfigure}%
    \begin{subfigure}{0.2\textwidth}
        \includegraphics[width=\textwidth]{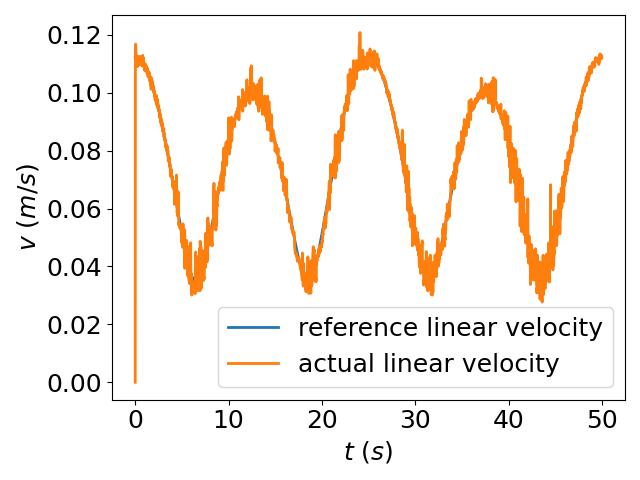}
        \caption{Linear velocity}
      
    \end{subfigure}%
    \begin{subfigure}{0.2\textwidth}
        \includegraphics[width=\textwidth]{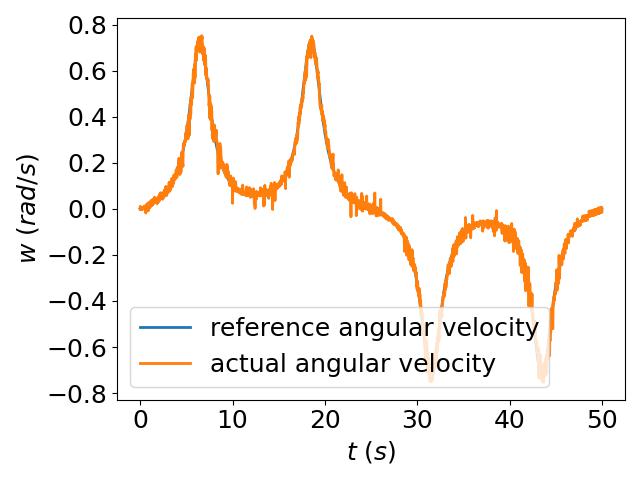}
        \caption{Angular velocity}
      
    \end{subfigure}%

    \caption{Butterfly-shaped trajectory tracking with Turtlebot 3. The initial position and initial orientation are aligned with the start position and the start orientation of the reference trajectory.}
    \label{fig:butterfly-res-turtle}
\end{figure*}

\begin{figure*}[tb]
    \centering
    \begin{subfigure}{0.2\textwidth}
        \includegraphics[width=\textwidth]{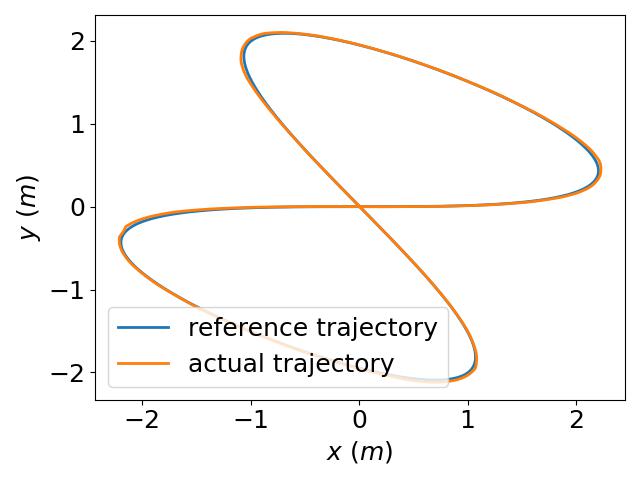}
        \caption{Trajectory (GMPC)}
        \label{fig:butterfly-gmpc-traj}
       
    \end{subfigure}
    \begin{subfigure}{0.2\textwidth}
        \includegraphics[width=\textwidth]{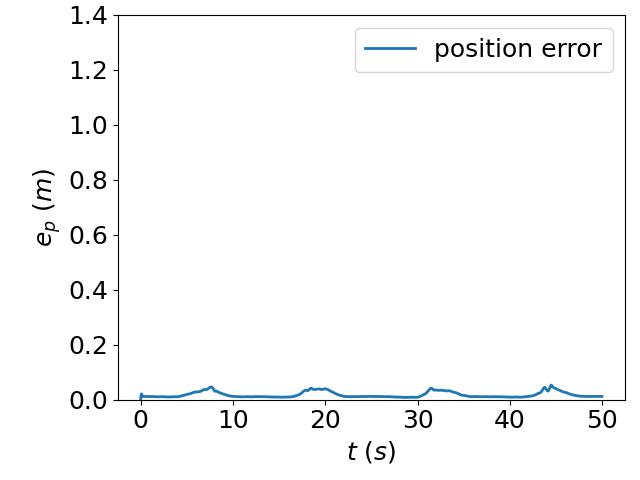}
        \caption{Position error}
       
    \end{subfigure}%
    \begin{subfigure}{0.2\textwidth}
        \includegraphics[width=\textwidth]{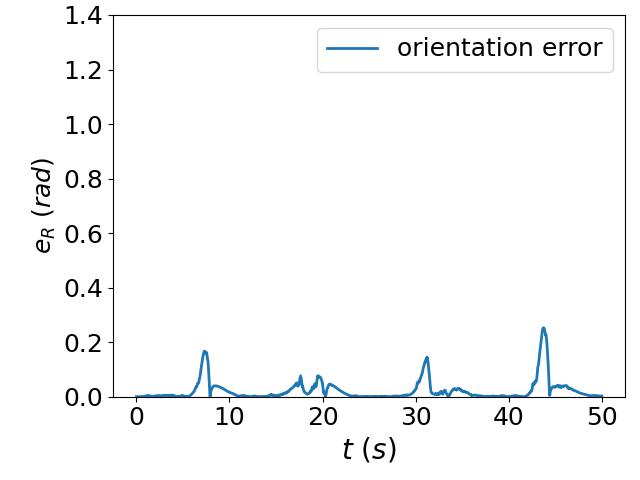}
        \caption{Orientation error}
      
    \end{subfigure}%
    \begin{subfigure}{0.2\textwidth}
        \includegraphics[width=\textwidth]{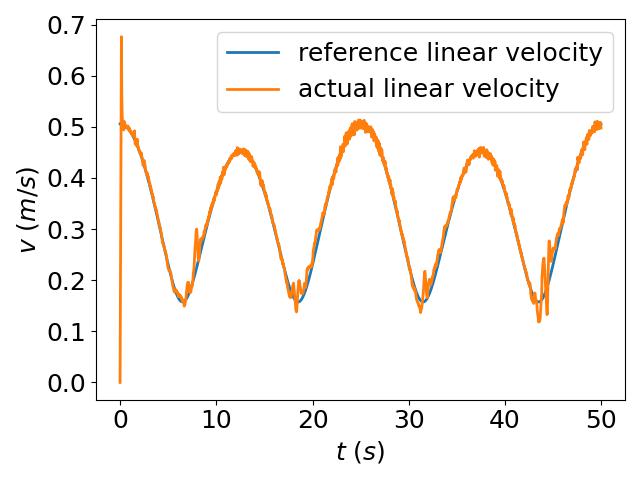}
        \caption{Linear velocity}
      
    \end{subfigure}%
    \begin{subfigure}{0.2\textwidth}
        \includegraphics[width=\textwidth]{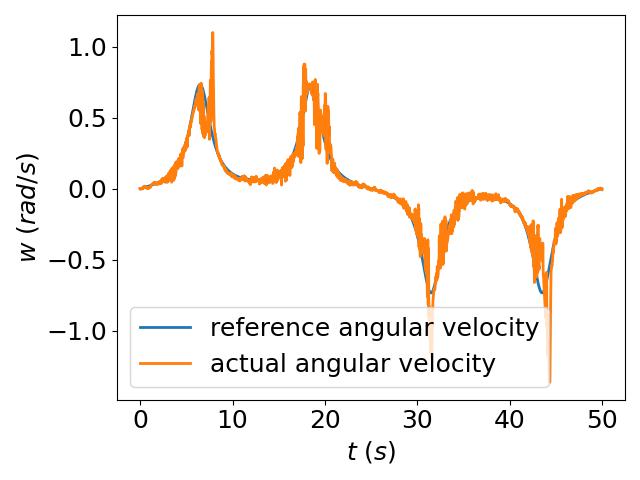}
        \caption{Angular velocity}
      
    \end{subfigure}%

    \quad
    \begin{subfigure}{0.2\textwidth}
        \includegraphics[width=\textwidth]{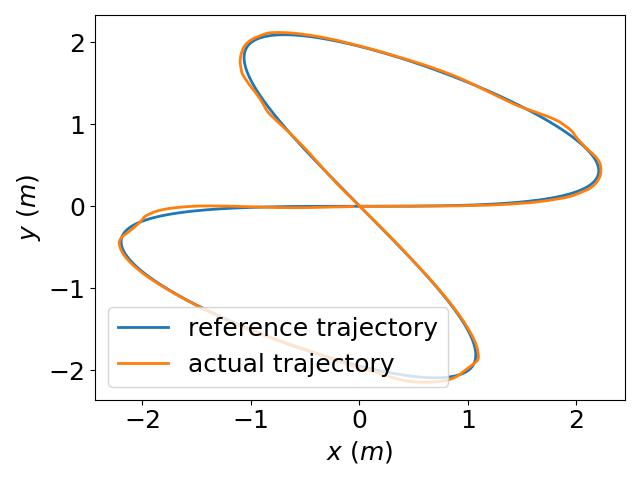}
        \caption{Trajectory (NMPC)}
        \label{fig:butterfly-nmpc-traj}
       
    \end{subfigure}
    \begin{subfigure}{0.2\textwidth}
        \includegraphics[width=\textwidth]{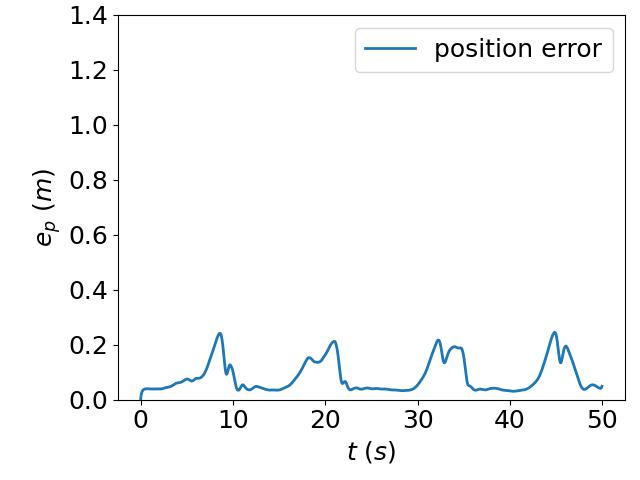}
        \caption{Position error}
         \label{fig:butterfly-nmpc-position-e}
       
    \end{subfigure}%
    \begin{subfigure}{0.2\textwidth}
        \includegraphics[width=\textwidth]{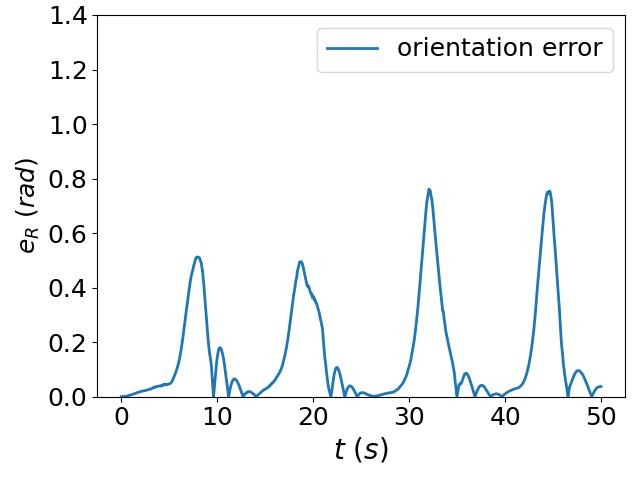}
        \caption{Orientation error}
         \label{fig:butterfly-nmpc-orientation-e}
      
    \end{subfigure}%
    \begin{subfigure}{0.2\textwidth}
        \includegraphics[width=\textwidth]{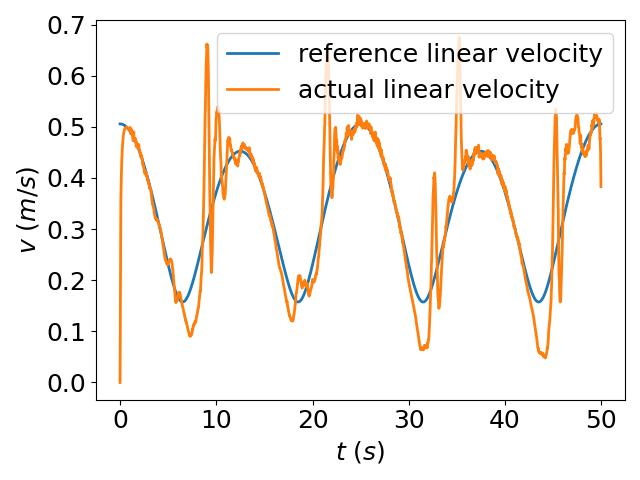}
        \caption{Linear velocity}
         \label{fig:butterfly-nmpc-linear-v}
      
    \end{subfigure}%
    \begin{subfigure}{0.2\textwidth}
        \includegraphics[width=\textwidth]{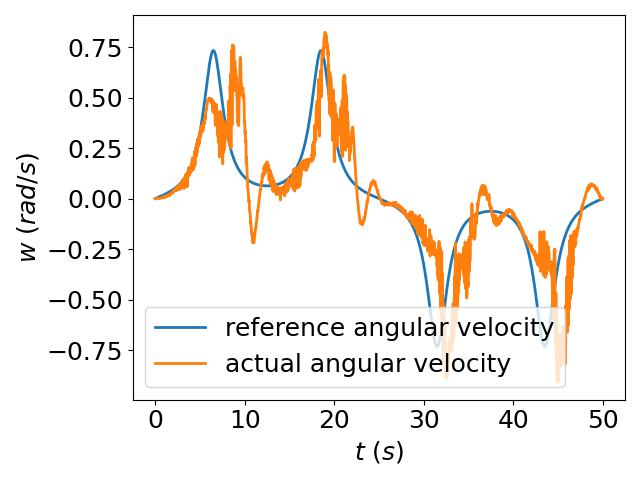}
        \caption{Angular velocity}
         \label{fig:butterfly-nmpc-angular-v}
      
    \end{subfigure}%

    \quad
    \begin{subfigure}{0.2\textwidth}
        \includegraphics[width=\textwidth]{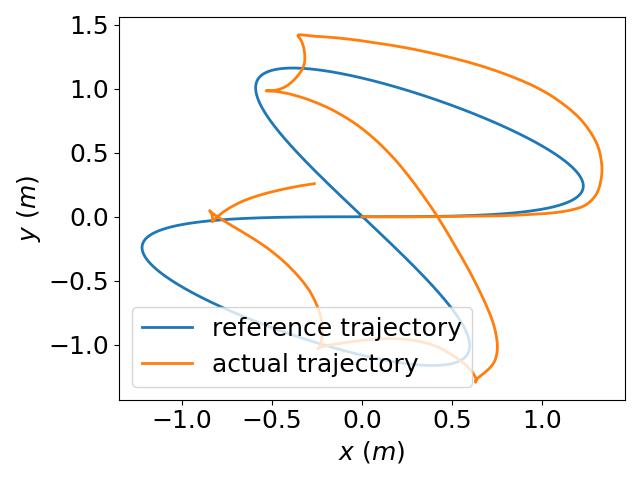}
        \caption{Trajectory (FBC)}
       
    \end{subfigure}
    \begin{subfigure}{0.2\textwidth}
        \includegraphics[width=\textwidth]{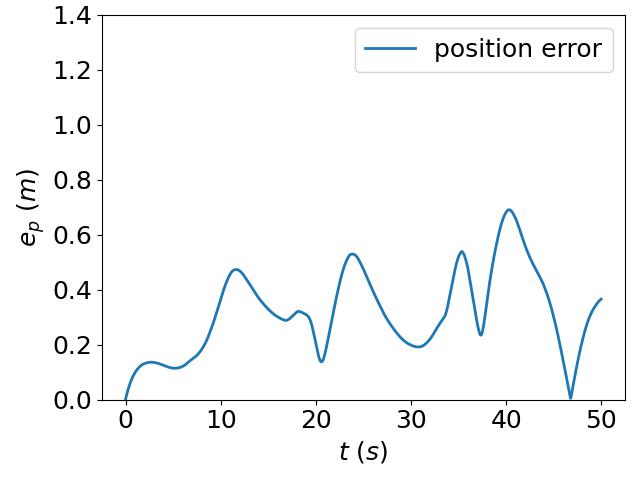}
        \caption{Position error}
       
    \end{subfigure}%
    \begin{subfigure}{0.2\textwidth}
        \includegraphics[width=\textwidth]{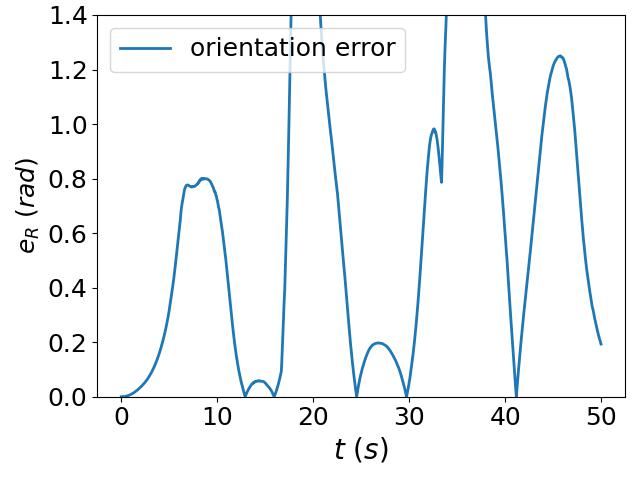}
        \caption{Orientation error}
      
    \end{subfigure}%
    \begin{subfigure}{0.2\textwidth}
        \includegraphics[width=\textwidth]{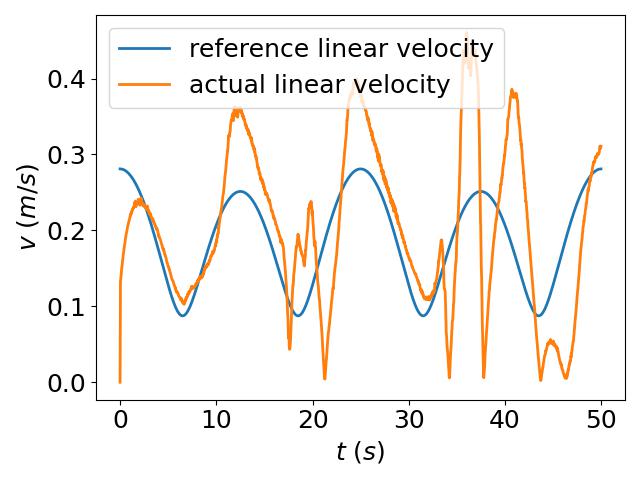}
        \caption{Linear velocity}
      
    \end{subfigure}%
    \begin{subfigure}{0.2\textwidth}
        \includegraphics[width=\textwidth]{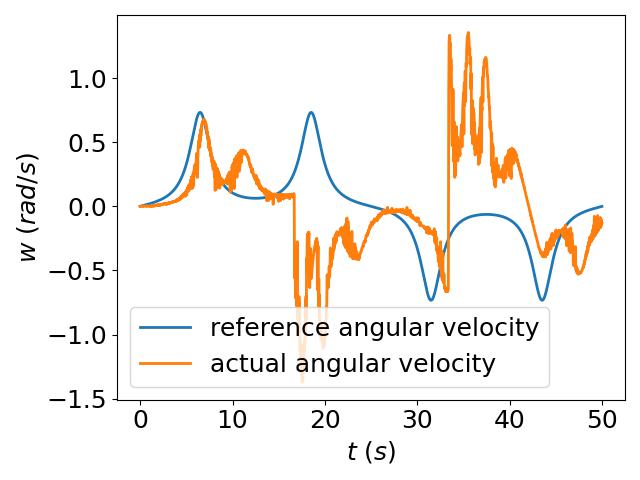}
        \caption{Angular velocity}
      
    \end{subfigure}%

    \caption{Butterfly-shaped trajectory tracking with Scout mini. The initial position and initial orientation are aligned with the start position and the start orientation of the reference trajectory.}
    \label{fig:butterfly-res}

\end{figure*}
\subsubsection{Linearization Schemes Comparison} In our GMPC implementation, we conduct a comparison of the linearization schemes discussed in Section III. The results of this comparison are shown in Fig. \ref{fig: linearization comparison}. It is evident from the figure that when employing the linearization scheme outlined in \eqref{linear error-dyn wrong}, the convergence of position error is relatively slow. {Furthermore, this particular linearization scheme leads to non-negligible steady-state errors when tracking trajectories.} In contrast, the linearization scheme described in \eqref{linear twist} exhibits notable superior performance. The position error convergence is significantly faster compared to the previous one. Furthermore, \eqref{linear twist} effectively mitigates the steady-state error issue that plagued the previous scheme, resulting in a much closer alignment between the actual and desired trajectory. 

\subsubsection{Circular Trajectory Tracking Comparison} Fig. \ref{fig:MC-TEST} shows the Monte Carlo test of tracking a circular trajectory with Turtlebot 3. In this scenario, three controllers demonstrate convergence in both position and orientation errors. Our controller outperforms other methods in terms of orientation convergence speed. \textcolor{black}{It can be easily verified that GMPC exhibits smooth convergence in terms of position error and orientation error by virtue of the continuity and smoothness properties of the matrix Lie group. Although NMPC can also successfully track circular trajectories, it encounters the singularity problem from Euler-based formulation and numerical instability from nonlinear equality constraints, resulting in overshoot and oscillation}. The FBC controller avoids overshooting but suffers from unstable feedback control due to the singularity issue when representing orientation with the Euler angle. Consequently, both $e_p(t)$ and $e_R(t)$ increase when the value of $\theta$ is around $\pi$ ($t = 16s$).

\subsubsection{Butterfly-shaped Trajectory Tracking Comparison} Fig. \ref{fig:butterfly-res-turtle} and Fig. \ref{fig:butterfly-res} display the results of tracking the butterfly-shaped trajectory using Turtlebot 3 and Scout mini, respectively. When employed on Turtlebot 3, all three controllers demonstrate excellent performance. However, when applied to Scout mini, it becomes evident that the FBC fails to track the trajectory successfully. The major reason is that using the same feedback gain, the FBC cannot adapt to the model residuals of the four-wheel-drive model. \textcolor{black}{GMPC can track the reference velocities closely and generate smooth actual trajectory. However, NMPC struggles to track the reference velocities closely (as shown by the actual linear and angular velocity shown in Fig. \ref{fig:butterfly-nmpc-linear-v} and Fig. \ref{fig:butterfly-nmpc-angular-v}), which results in large position error and orientation error and unsmooth actual trajectory. (as shown in Fig. \ref{fig:butterfly-nmpc-position-e}, Fig. \ref{fig:butterfly-nmpc-orientation-e} and Fig. \ref{fig:butterfly-gmpc-traj}). The improvement of our GMPC showcased the advantage of the matrix Lie group formulation.}
\subsection{Physical Experiments}
\begin{figure*}[htb]
    \centering
    \begin{subfigure}{0.33\textwidth}
        \includegraphics[width=\textwidth]{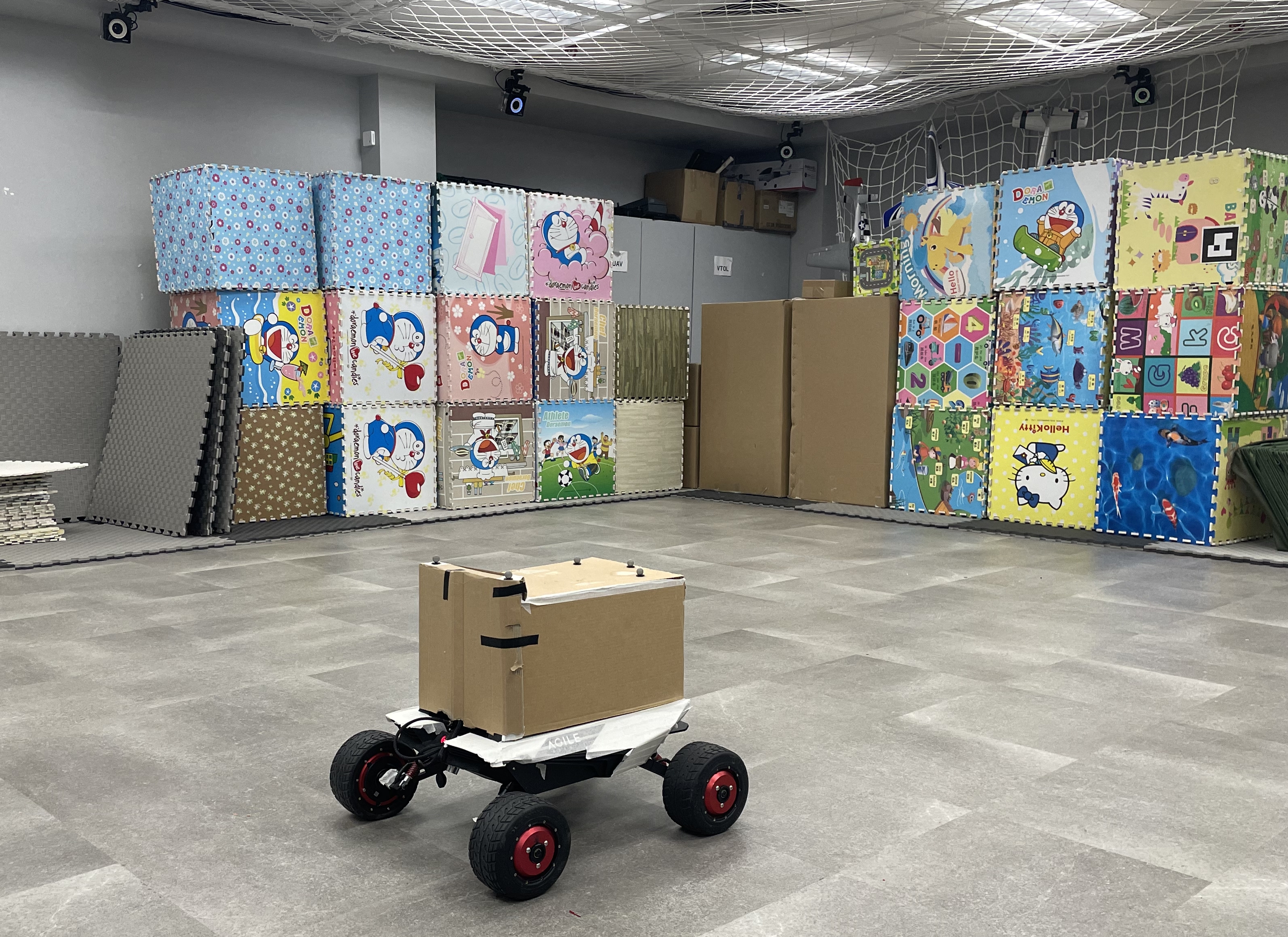}
        \caption{\textcolor{black}{Physical experiment arena}}
        \label{fig:physical_arena}
    \end{subfigure}
    \begin{subfigure}{0.33\textwidth}
        \includegraphics[width=\textwidth]{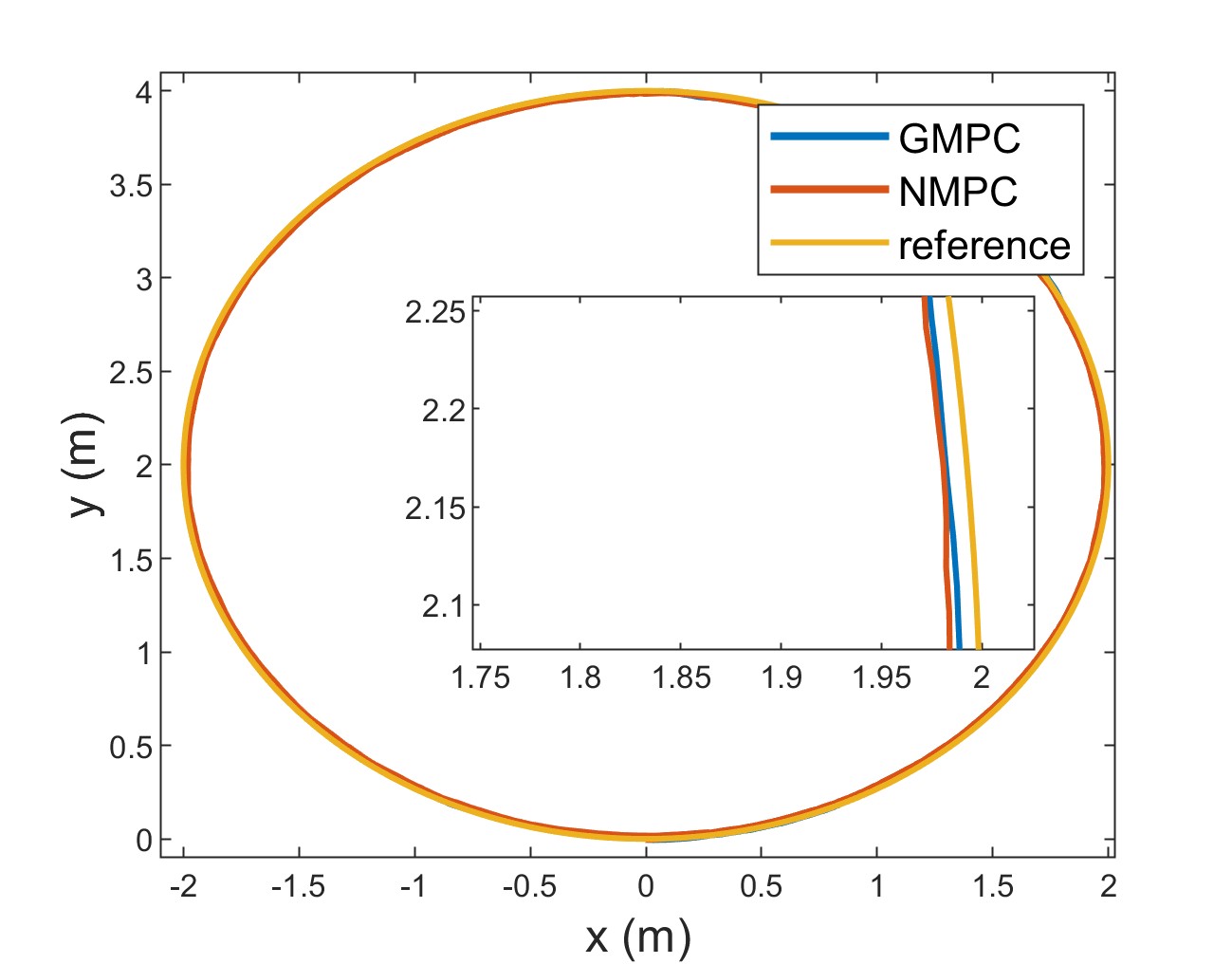}
        \caption{\textcolor{black}{Physical experiment trajectories comparison}}
        \label{fig:physical_traj}
    \end{subfigure}%
    \begin{subfigure}{0.33\textwidth}
        \includegraphics[width=\textwidth]{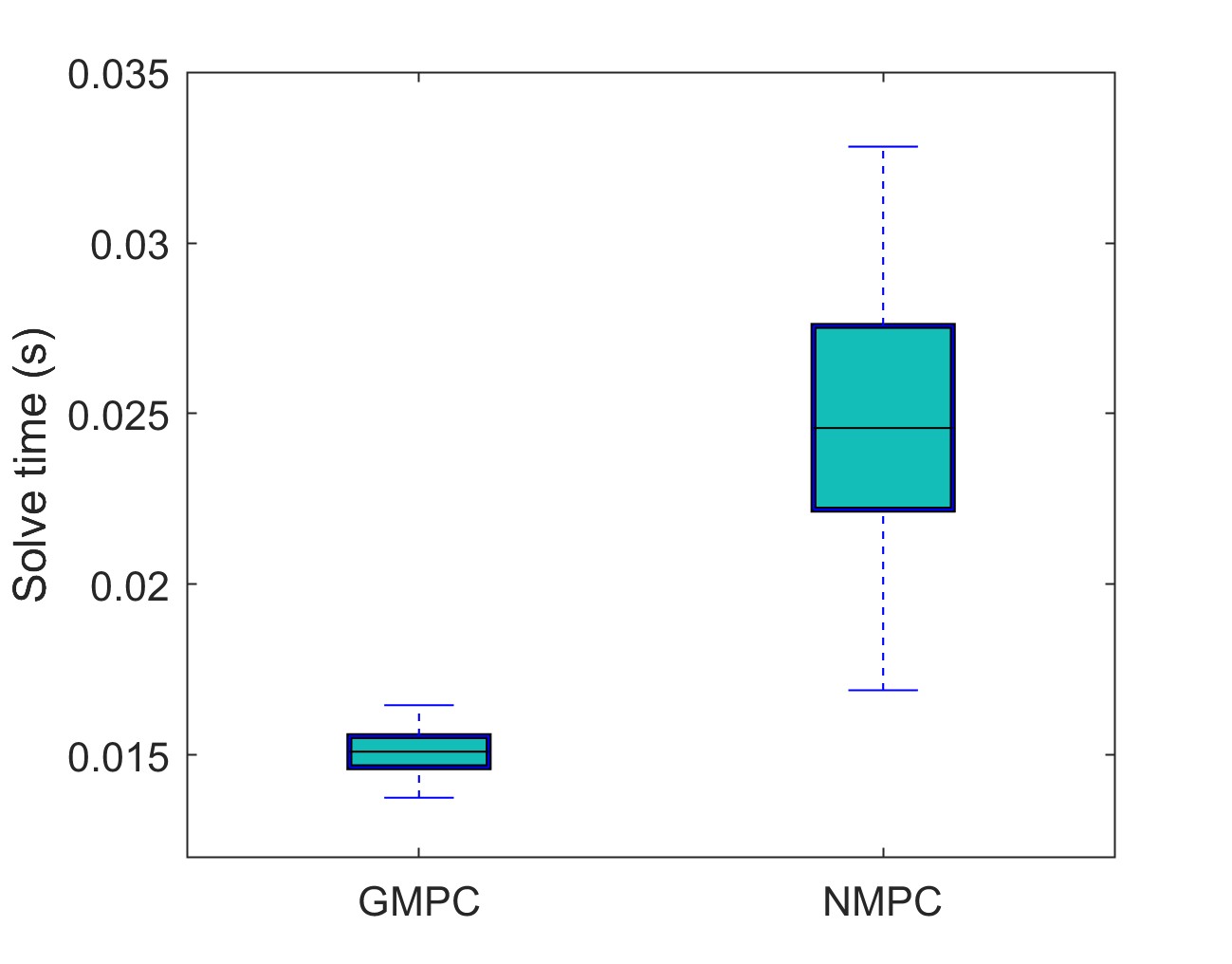}
        \caption{Physical experiment solve time comparison}
        \label{fig:solve_time}
    \end{subfigure}%
   
    \caption{Physical experiment arena, trajectories comparison and solve time comparison.}
\end{figure*}

In our physical experiment, we use a Scout mini equipped with Intel NUC 11 (Intel i7-1165G7 CPU) as our platform. The experiment takes place indoors, and the OptiTrack Motion Capture system, consisting of 14 cameras, estimates the state of the WMR. \textcolor{black}{Fig. \ref{fig:physical_arena} shows our physical experiment arena.} The task for the WMR is to track the circular trajectory with a control frequency of 5 Hz. As shown in Fig. \ref{fig:physical_traj}, both GMPC and NMPC can track the reference trajectory \textcolor{black}{with a 1cm position error.} However, notable differences were observed in their runtime performance. The quarterback chart comparison in Fig. \ref{fig:solve_time} demonstrates that our GMPC outperforms NMPC in terms of solving time, being approximately 1.5 times faster. Furthermore, the GMPC solving time variance is significantly lower than NMPC. The rationale behind the improved runtime performance of our GMPC controller is its convex formulation for trajectory tracking. As we convert the problem to the vector space and formulate the convex optimization problem, we can adopt QP solvers to solve the problem. This offers computational advantages in efficiency and stability over solving nonlinear programming with NLP solvers.

\section{Conclusion}
In this letter, we presented a novel geometric model predictive control framework for wheeled mobile robot trajectory tracking. We explored the relationship between the Lie group and the corresponding Lie algebra of the WMR to derive its error dynamics for trajectory tracking. Through choosing a suitable linearization scheme, we converted the problem to vector space and formulated the convex optimal control problem, which can be solved efficiently with QP solvers. The simulations and physical experiments demonstrated the superior performance of the proposed method. Future research directions include developing a high-quality trajectory generation algorithm and building a unified planning and control pipeline for WMRs. We hope that our simulator will make it easier for researchers from robotic communities to develop advanced control algorithms for different robotic systems.


\bibliographystyle{IEEEtran}
\bibliography{reference}

\begin{thebibliography}{10}
\providecommand{\url}[1]{#1}
\csname url@samestyle\endcsname
\providecommand{\newblock}{\relax}
\providecommand{\bibinfo}[2]{#2}
\providecommand{\BIBentrySTDinterwordspacing}{\spaceskip=0pt\relax}
\providecommand{\BIBentryALTinterwordstretchfactor}{4}
\providecommand{\BIBentryALTinterwordspacing}{\spaceskip=\fontdimen2\font plus
\BIBentryALTinterwordstretchfactor\fontdimen3\font minus \fontdimen4\font\relax}
\providecommand{\BIBforeignlanguage}[2]{{%
\expandafter\ifx\csname l@#1\endcsname\relax
\typeout{** WARNING: IEEEtran.bst: No hyphenation pattern has been}%
\typeout{** loaded for the language `#1'. Using the pattern for}%
\typeout{** the default language instead.}%
\else
\language=\csname l@#1\endcsname
\fi
#2}}
\providecommand{\BIBdecl}{\relax}
\BIBdecl

\bibitem{ma-tits-2023}
Z.~Lin, J.~Ma, J.~Duan, S.~E. Li, H.~Ma, B.~Cheng, and T.~H. Lee, ``Policy iteration based approximate dynamic programming toward autonomous driving in constrained dynamic environment,'' \emph{IEEE Transactions on Intelligent Transportation Systems}, vol.~24, no.~5, pp. 5003--5013, 2023.

\bibitem{warehouse-ral}
Z.~He, X.~Zhang, S.~Jones, S.~Hauert, D.~Zhang, and N.~F. Lepora, ``{TacMMs}: Tactile mobile manipulators for warehouse automation,'' \emph{IEEE Robotics and Automation Letters}, vol.~8, no.~8, pp. 4729--4736, 2023.

\bibitem{caochao-sr}
C.~Cao, H.~Zhu, Z.~Ren, H.~Choset, and J.~Zhang, ``Representation granularity enables time-efficient autonomous exploration in large, complex worlds,'' \emph{Science Robotics}, vol.~8, no.~80, 2023.

\bibitem{261398}
C.~Canudas~de Wit and O.~Sordalen, ``Exponential stabilization of mobile robots with nonholonomic constraints,'' in \emph{the 30th IEEE Conference on Decision and Control}, 1991, pp. 692--697 vol.1.

\bibitem{ASTOLFI199637}
A.~Astolfi, ``Discontinuous control of nonholonomic systems,'' \emph{Systems \& Control Letters}, vol.~27, no.~1, pp. 37--45, 1996.

\bibitem{5400088}
D.~Kostić, S.~Adinandra, J.~Caarls, N.~van~de Wouw, and H.~Nijmeijer, ``Collision-free tracking control of unicycle mobile robots,'' in \emph{Proceedings of the 48h IEEE Conference on Decision and Control (CDC) held jointly with 28th Chinese Control Conference}, 2009, pp. 5667--5672.

\bibitem{fBL}
G.~Oriolo, A.~De~Luca, and M.~Vendittelli, ``{WMR} control via dynamic feedback linearization: design, implementation, and experimental validation,'' \emph{IEEE Transactions on Control Systems Technology}, vol.~10, no.~6, pp. 835--852, 2002.

\bibitem{4812087}
P.~Morin and C.~Samson, ``Control of nonholonomic mobile robots based on the transverse function approach,'' \emph{IEEE Transactions on Robotics}, vol.~25, no.~5, pp. 1058--1073, 2009.

\bibitem{PLIEGOJIMENEZ2021109756}
J.~Pliego-Jiménez, R.~Martínez-Clark, C.~Cruz-Hernández, and A.~Arellano-Delgado, ``Trajectory tracking of wheeled mobile robots using only cartesian position measurements,'' \emph{Automatica}, vol. 133, p. 109756, 2021.

\bibitem{9981038}
Z.~Jian, Z.~Lu, X.~Zhou, B.~Lan, A.~Xiao, X.~Wang, and B.~Liang, ``{PUTN}: A plane-fitting based uneven terrain navigation framework,'' in \emph{IEEE/RSJ International Conference on Intelligent Robots and Systems (IROS)}, 2022, pp. 7160--7166.

\bibitem{10160857}
Z.~Jian, Z.~Yan, X.~Lei, Z.~Lu, B.~Lan, X.~Wang, and B.~Liang, ``Dynamic control barrier function-based model predictive control to safety-critical obstacle-avoidance of mobile robot,'' in \emph{IEEE International Conference on Robotics and Automation (ICRA)}, 2023, pp. 3679--3685.

\bibitem{10036044}
J.~Song, G.~Tao, Z.~Zang, H.~Dong, B.~Wang, and J.~Gong, ``Isolating trajectory tracking from motion control: A model predictive control and robust control framework for unmanned ground vehicles,'' \emph{IEEE Robotics and Automation Letters}, vol.~8, no.~3, pp. 1699--1706, 2023.

\bibitem{KHAN2022103903}
S.~Khan, J.~Guivant, and X.~Li, ``Design and experimental validation of a robust model predictive control for the optimal trajectory tracking of a small-scale autonomous bulldozer,'' \emph{Robotics and Autonomous Systems}, vol. 147, p. 103903, 2022.

\bibitem{8967788}
T.~A. Howell, B.~E. Jackson, and Z.~Manchester, ``{ALTRO}: A fast solver for constrained trajectory optimization,'' in \emph{IEEE/RSJ International Conference on Intelligent Robots and Systems (IROS)}, 2019, pp. 7674--7679.

\bibitem{micro-lie-theory}
J.~Sol{\`a}, J.~Deray, and D.~Atchuthan, ``A micro lie theory for state estimation in robotics,'' \emph{ArXiv}, vol. abs/1812.01537, 2018.

\bibitem{PWA}
B.~E. Jackson, K.~Tracy, and Z.~Manchester, ``Planning with attitude,'' \emph{IEEE Robotics and Automation Letters}, vol.~6, no.~3, pp. 5658--5664, 2021.

\bibitem{Alcan2023TrajectoryOO}
G.~Alcan, F.~J. Abu-Dakka, and V.~Kyrki, ``Trajectory optimization on matrix lie groups with differential dynamic programming and nonlinear constraints,'' \emph{ArXiv}, vol. abs/2301.02018, 2023.

\bibitem{ommpc}
G.~Lu, W.~Xu, and F.~Zhang, ``On-manifold model predictive control for trajectory tracking on robotic systems,'' \emph{IEEE Transactions on Industrial Electronics}, vol.~70, no.~9, pp. 9192--9202, 2023.

\bibitem{9981282}
S.~Teng, D.~Chen, W.~Clark, and M.~Ghaffari, ``An error-state model predictive control on connected matrix lie groups for legged robot control,'' in \emph{IEEE/RSJ International Conference on Intelligent Robots and Systems (IROS)}, 2022, pp. 8850--8857.

\bibitem{iekf}
A.~Barrau and S.~Bonnabel, ``The invariant extended kalman filter as a stable observer,'' \emph{IEEE Transactions on Automatic Control}, vol.~62, no.~4, pp. 1797--1812, 2017.

\bibitem{betts2010practical}
J.~T. Betts, \emph{{Practical Methods for Optimal Control and Estimation Using Nonlinear Programming}}.\hskip 1em plus 0.5em minus 0.4em\relax SIAM, 2010.

\bibitem{osqp}
B.~Stellato, G.~Banjac, P.~Goulart, A.~Bemporad, and S.~Boyd, ``{OSQP}: an operator splitting solver for quadratic programs,'' \emph{Mathematical Programming Computation}, vol.~12, no.~4, pp. 637--672, 2020.

\bibitem{Ferreau2014}
H.~Ferreau, C.~Kirches, A.~Potschka, H.~Bock, and M.~Diehl, ``{qpOASES}: A parametric active-set algorithm for quadratic programming,'' \emph{Mathematical Programming Computation}, vol.~6, no.~4, pp. 327--363, 2014.

\bibitem{pybullet}
E.~Coumans and Y.~Bai, ``{PyBullet, a Python module for physics simulation for games, robotics and machine learning},'' \url{http://pybullet.org}, 2016--2021.

\bibitem{andersson2019casadi}
J.~A. Andersson, J.~Gillis, G.~Horn, J.~B. Rawlings, and M.~Diehl, ``{CasADi}: a software framework for nonlinear optimization and optimal control,'' \emph{Mathematical Programming Computation}, vol.~11, pp. 1--36, 2019.

\bibitem{Manif}
J.~Deray and J.~Solà, ``Manif: A micro {L}ie theory library for state estimation in robotics applications,'' \emph{Journal of Open Source Software}, vol.~5, no.~46, p. 1371, 2020.

\bibitem{wachter2006implementation}
A.~W{\"a}chter and L.~T. Biegler, ``On the implementation of an interior-point filter line-search algorithm for large-scale nonlinear programming,'' \emph{Mathematical Programming}, vol. 106, pp. 25--57, 2006.

\bibitem{Lee2019IntroductionTR}
J.~M. Lee, \emph{Introduction to Riemannian Manifolds}.\hskip 1em plus 0.5em minus 0.4em\relax Springer, 2019.

\end{thebibliography}
\end{document}